\documentclass[preprints,aricle,accept,pdftex,moreauthors]{Definitions/mdpi} 
\firstpage{1} 
\pubvolume{1}
\issuenum{1}
\articlenumber{0}
\pubyear{2022}
\copyrightyear{2022}
\datereceived{} 
\dateaccepted{} 
\datepublished{} 
\hreflink{https://doi.org/} 

\usepackage{amssymb}
\usepackage{subcaption}
\usepackage{textcomp}

\newcommand*\diff{\mathop{}\!\mathrm{d}}

\Title{\textit{In silico} evaluation of Paxlovid's pharmacometrics for SARS--CoV--2: a multiscale approach}

\TitleCitation{\textit{In silico} evaluation of Paxlovid's pharmacometrics for SARS--CoV--2: a multiscale approach}

\Author{Ferenc A. Bartha $^{1}$\orcidA{}, N\'ora Juh\'asz $^{1}$\orcidB{}, Sadegh Marzban $^1$\orcidC{}, Renji Han $^{2}$\orcidD{} and Gergely R\"ost$^{1}$\orcidE{}}

\AuthorNames{Ferenc A. Bartha, N\'ora Juh\'asz, Sadegh Marzban, Renji Han and Gergely R\"ost}

\AuthorCitation{Bartha, F.A.; Juh\'asz, N.; Marzban, S.; Han, R; R\"ost, G.}

\address{%
$^{1}$ \quad Bolyai Institute, University of Szeged, H-6720 Szeged, Hungary; barfer@math.u-szeged.hu (F.A.B.); juhaszn@math.u-szeged.hu (N.J.); sadegh.marzban@math.u-szeged.hu (S.M.); rost@math.u-szeged.hu (G.R.)\\
$^{2}$ \quad Zhejiang University of Science and Technology, 310023 Hangzhou Zhejiang, China; renjihan@csu.edu.cn (R.H.)}

\corres{Correspondence: barfer@math.u-szeged.hu (F.A.B.); juhaszn@math.u-szeged.hu (N.J.)}

\abstract{Paxlovid is a promising, orally bioavailable novel drug for SARS--CoV--2 with excellent safety profiles. Our main goal here is to explore the pharmacometric features of this new antiviral. To provide a detailed assessment of Paxlovid, we propose a hybrid multiscale mathematical approach. We demonstrate that the results of the present \textit{in silico} evaluation match the clinical expectations remarkably well: on the one hand, our computations successfully replicate the outcome of an actual \textit{in vitro} experiment; on the other hand we verify both the sufficiency and the necessity of Paxlovid's two main components (nirmatrelvir and ritonavir) for a simplified \textit{in vivo} case. Moreover, in the simulated context of our computational framework we visualize the importance of early interventions, and identify the time window where a unit--length delay causes the highest level of tissue damage. Finally, the results' sensitivity to the diffusion coefficient of the virus is explored in details.}


\begin{document}

\section{Introduction}

Even with extensive vaccination, COVID--19 will most probably not be eradicated from the human populations, thus new options for therapy need to be explored. This paper concentrates on the mathematical evaluation, assessment, and computation-based simulation of a promising antiviral drug, Paxlovid \cite{nirmatrelvir-ritonavir}, which is essentially nirmatrelvir co-packaged with ritonavir. Nirmatrelvir is a protease inhibitor that is active against $\text{M}^\text{pro}$: inhibition of the SARS--CoV--2 main protease renders it incapable of processing polyprotein precursors, preventing virus production. Ritonavir is given as a pharmacokinetic enhancer: it slows down nirmatrelvir’s metabolism allowing a twice daily administration regimen. Paxlovid (also known as PF-07321332) is orally bioavailable and it has excellent \textit{in vivo} safety profiles \cite{Mpro-inhibitor, fda-fact-sheet, ema}.

Modeling has enhanced our understanding of the dynamics of viral spread, and it played an instrumental role in developing successful therapies for chronic viral infections such as HIV and HCV \cite{Perelson-Ke}. Mathematical models have proved to be indispensable tools in overcoming the challenges posed by the SARS--CoV--2 pandemic, too. In terms of investigating cellular--level antiviral dynamics, one of the most modern, state-of-the-art approaches consists in considering each component's physical dimension and capturing them within the framework of a spatial multiscale model accordingly. These systems incorporate size in a particular manner: rather than operating with a simple numerical value, what change in these models -- depending on the size of a given biological entity -- are the mathematical tools themselves that are applied to grasp these variables on different scales.

More formally, building on our previous work \cite{hybrid-PDE-ABM-1}, we define a hybrid mathematical model by merging \textit{i)} a partial differential equation representing local virus concentration, \textit{ii)} an agent-based model describing target cells in the lung and their three possible states (uninfected, infected, and dead), and \textit{iii)} a partial differential equation representing nirmatrelvir concentration. Naturally, the respective parts are closely and meaningfully intertwined: each considerable interaction and feedback process connecting these separate biological participants is given formal definition and appears in the implementation. A detailed motivation, definition and construction of this model type and its various advantages are discussed in \cite{hybrid-PDE-ABM-1} -- here we limit ourselves to highlighting the role of including crucial spatial mechanisms: unlike other classical models such as the ODE approach, the present system is defined in both space and time, and consequently is able to capture highly significant -- and in nature inherently spatial -- physical phenomena such as virus diffusion.
 
Our goal here is to provide an assessment of Paxlovid based on mathematical simulations. We explore questions such as what happens if nirmatrelvir is taken without ritonavir, or how do expectations for treatment outcome change if tablets are taken with some delay. We emphasise that performing the analogous \textit{in vivo} experiments on an actual person or animal would be either simply impossible or unethical, at the same time the corresponding \textit{in silico} experiment can be conducted in just a few minutes at an ideally low cost.

While various other software packages implement hybrid mathematical concepts \cite{compucell, physicell}, we highlight that our implementation of the proposed multiscale system modeling Paxlovid is based on an adaptation of the free and open source library, HAL (Hybrid Automata Library) \cite{Bravo2020}.

\section{Methods}

\subsection{The hybrid PDE-ABM model}

As described in the Introduction, the main multiscale framework is defined via forming meaningful bridges between two important and fundamentally different modeling techniques: we merge continuous partial differential equations and a discrete agent based model.

We begin by setting a notation: let $\Omega$ be the mathematical representation of the area we are considering. For an \textit{in vivo} experiment this would mean a small part of the lung tissue, while in the case of an \textit{in vitro} experiment it would be the area of a single well in a laboratory plate. Now we are ready to introduce \textbf{the discrete part} of our hybrid model.

\subsubsection{Epithelial cells}

One of the most important modeling decisions in \cite{hybrid-PDE-ABM-1} was defining epithelial cells as discrete agents. The diameter of epithelial cells is relatively significant \cite{alveolartypeIIcovid, alvtype2cellsize} and consequently, in terms of mathematical conceptualization it is natural to approach cells as separate entities and follow their respective states on an individual level.

The discrete state space of target cells is defined precisely as in \cite{hybrid-PDE-ABM-1}. For completeness, we recall some of the fundamental technical details; namely, we construct a two dimensional ABM state space by introducing a lattice of $k_1 \times k_2$ agents representing epithelial cells ($k_1, k_2 \in \mathbb{N}$). Cells are identified by means of the corresponding agent's place in the grid, or formally, by the $(i,j)$ indices, where $(i,j) \in {\mathcal{J}} = \{(i,j) \vert 1 \leq i \leq k_1,$ $1 \leq j \leq k_2 \}$. Finally, by setting the $\Omega_{i,j}$ notation for the open set occupied by the $(i,j)$-th cell, we have $\bar{\Omega} = \bigcup\limits_{(i,j)\in {\mathcal{J}}} \bar{\Omega}_{i,j}$.

Regarding cell states in the context of the ABM space, the main concept is rather straightforward: each agent has three potential states. The latter is formally captured by following $s_{i,j}(t)$ state function, which represents the basic idea that an epithelial lung cell is either \textit{uninfected}, \textit{infected}, or \textit{dead}:

\begin{linenomath}
\begin{equation*}
  s_{i,j}(t) =
  \begin{cases}
  	\text{T}, & \text{if the $(i,j)$-th cell is alive and uninfected at time $t$} \\
    \text{I}, & \text{if the $(i,j)$-th cell is infected at time $t$} \\
  	\text{D}, & \text{if the $(i,j)$-th cell is dead at time $t$.}
  \end{cases}
\end{equation*}
\end{linenomath}

We note that the \textit{uninfected}, \textit{susceptible}, and \textit{target (cell)} expressions are used interchangeably in the context of viral dynamics: they all refer to living cells that are susceptible to SARS--CoV--2 infection but are (for the time being) free from it.

Concerning state dynamics, the transition rules are set to naturally mimic the biological phenomenon in question, the complete list is as follows:

\begin{itemize}
    \item all living uninfected cells are susceptible target cells to virus infection;
    \item since the time frame of infection is relatively short, cell birth and cell division are ignored;
    \item infection is not reversible: an infected cell can not become a healthily functioning uninfected cell again;
    \item viral infection itself is the only reason for cell death, i.e. death related to any other natural cause is not accounted for;
    \item the \textit{uninfected} $\to$ \textit{infected} state change: a target cell may become infected depending on the local virus concentration at the given cell. Infection itself is randomized and it occurs with a probability of $P_I$ (for more details see \cite{hybrid-PDE-ABM-1});
    \item the \textit{infected} $\to$ \textit{dead} state change: an infected cell dies with a probability of $P_D.$ Analogously to infection, death is approached from a stochastic viewpoint as well.
\end{itemize}

The cornerstone observation motivating the modeling decision behind the second, \textbf{continuous part} of the framework is that viruses and drug molecules are several magnitudes smaller than epithelial cells \cite{virussize} -- we incorporate this simple but crucial biological feature into our system by modeling both virus concentration $V$ and drug concentration $N$ as continuous functions.

\subsubsection{Virus concentration}

As suggested above, virus concentration $V(t,x,y)$ is described as a variable that is continuously changing in both space and time, and as such, it is formally described by means of a PDE:

\begin{linenomath}
\begin{equation}
\label{virus-concentration}
    \left\{
    \begin{array}{llll}
        
        \frac{\partial V(t,x,y)}{\partial t}= D_V \Delta V - \mu_V V + (1 - \eta_N(N)) \cdot \sum\limits_{(i,j) \in \mathcal{J}} g_{i,j}(t,x,y), \quad t>0, \,\, (x,y) \in \Omega, \\\\
 
        \frac{\partial V(t,x,y)}{\partial\nu} = 0, \quad t>0, \,\, (x,y)\in \partial \Omega,
  
    \end{array}
    \right.
\end{equation}
\end{linenomath}

where $D_V$ stands for the virus diffusion coefficient, $\mu_V$ represents the virus removal constant, $N$ is the local concentration of nirmatrelvir (i.e. the active antiviral component of Paxlovid), $\eta_N$ is the efficacy function of nirmatrelvir, while $g_{i,j}$ denotes the viral source term for the $(i,j)$-th cell.

Equation \eqref{virus-concentration} formulates the following basic ideas.

\begin{enumerate}[label=\roman*)]

    \item Virus particles spread across the domain primarily via diffusion.
    \item A non-specific, non-adaptive, simplified immune system is assumed which removes viruses in a constant ratio.
    \item A local nirmatrelvir concentration of $N(t,x,y)$ reduces virus production from infected cells by a ratio of $\eta_N(N(t,x,y)),$ for more details see Section~\ref{drug-concentration-section}.
    \item Infected cells generate new virus particles in a process that is formally described by the $g_{i,j}$ source functions:
    
    \begin{linenomath}
    \begin{equation}
    \label{formalgdef}
    g_{i,j} (t,x,y) =
    \begin{cases}
        0, & \text{if $s_{i,j}(t) = \text{T}$ and $(x,y) \in \Omega_{i,j}$} \\
        f_{i,j}(t,x,y), & \text{if $s_{i,j}(t) = \text{I}$ and $(x,y) \in \Omega_{i,j}$} \\
        0, & \text{if $s_{i,j}(t) = \text{D}$ and $(x,y) \in \Omega_{i,j}$} \\
        0 & \text{if $(x,y) \notin \Omega_{i,j}$}
    \end{cases}
    \end{equation}
    \end{linenomath}
    
    Roughly speaking, in general any reasonable $f_{i,j}(t,x,y)$ function is allowed in the above formula (for more details see \cite{hybrid-PDE-ABM-1} and \cite{Beauchemin2006}). In our case we worked with a constant setting using the estimate $f_{i,j} = 3.72 \cdot 10^{-3}$ copies / (ml $\cdot$ minute $\cdot$ cell) obtained by \cite{fijvalue}.

\end{enumerate}

\subsubsection{Drug concentration}\label{drug-concentration-section}

As briefly discussed in the Introduction, Paxlovid (also known as PF-07321332) is an an orally administered SARS--CoV--2 main protease inhibitor \cite{Mpro-inhibitor, fda-fact-sheet, ema}. Paxlovid is essentially a combination of two different drugs: nirmatrelvir -- capable of effectively blocking virus production in infected cells -- acts as its main antiviral component, while ritonavir serves to slow down the metabolism of nirmatrelvir to maintain significantly higher concentrations of the participant responsible for $\text{M}^\text{pro}$-inhibition. We emphasise that nirmatrelvir and ritonavir are not only separate entities as acting components: the corresponding drugs themselves are packaged in individual, separate tablets -- this means that the theoretical possibility to take, for example, nirmatrelvir only (without the beneficial effect of ritonavir) is very much given. This subsection is dedicated to formulate these statements in the context of the mathematical framework.

Based on the formerly detailed principles regarding the dimension of each participant, we naturally define nirmatrelvir concentration as a continuous variable and denote it by $N(t,x,y).$ We highlight that -- both for simplicity and because of the apparent lack of clinical data -- we do not explicitly introduce the analogous $R(t,x,y)$ function for ritonavir concentration. Instead, we focus only on two specific cases: ritonavir is either taken as instructed (i.e. $100$ mg of ritonavir every 12 hours), or not taken at all. Formally, we introduce the boolean $r$ to mathematically grasp the above concept:

\begin{linenomath}
\begin{equation*}
  r =
  \begin{cases}
  	\text{true}, & \text{if ritonavir is taken following official regimen,} \\
  	\text{false}, & \text{if ritonavir is not administered at all.}
  \end{cases}
\end{equation*}
\end{linenomath}

Practically, instead of operating with an $R(t,x,y)$ function, the model's information regarding ritonavir becomes a boolean in the background, indirectly controlling $N$ through the metabolism-related descriptors.

Two further anatomical details need to be taken into account before we can formulate the equations for drug concentration.

Firstly, capillary density is very high in the lung. In more detail, for the case of rats, there are about $11$ epithelial cells per a single alveolus \cite{encyclopediaRespMed}. Considering that there are approximately $40$ capillary loops per alveolus \cite{estimationAlveolarCapillaries}, this gives cca. $4$ capillary loops per epithelial cell. Consequently, because of the abundant presence of neighbouring capillaries for a single cell, it is not unnatural to assume a completely homogeneous drug distribution in the alveolar epithelium. As a result, we work with $N(t)$ instead of $N(t,x,y),$ and the equation describing nirmatrelvir concentration becomes an ODE instead of a PDE.

Secondly, in order to reproduce the characteristic local concentration curves observed in clinical data (we refer to Figure 2A in \cite{nirmatrelvir-decay-data}), we apply a standard pharmacokinetic two-compartment approach (for example \cite{2compartment-example} used a dual compartment PK model assessing antiviral therapy targeting SARS--CoV--2). The latter model consists of a central compartment (e.g. stomach) responsible for first-level drug metabolism and a peripheral one (in our case the lung) containing the target site of nirmatrelvir. Technically, we introduce an additional $c(t)$ function representing drug concentration at the central compartment -- this is the the amount of nirmatrelvir that is already present in the patient's system, but is not yet locally available at the level of the lung's epithelial cells. In the context of the two-compartment model, $N(t)$ corresponds to the peripheral compartment's nirmatrelvir concentration. Similarly to the case of $N(t),$ we use an ODE to describe $c(t)$.

The complete system for nirmatrelvir concentration is formally described by the following set of equations:
    
\begin{linenomath}
\begin{equation}
\label{drug-concentration}
    \left\{
    \begin{array}{llll}
        
        \frac{\diff c(t)}{\diff t} = - \mu_c(r) c(t) + S(t,r), \\\\
  
        \frac{\diff N(t)}{\diff t} =  - \mu_N(r) N(t) + \mu_c(r) c(t).
  
    \end{array}
    \right.
\end{equation}
\end{linenomath}
    
In the above set of equations $S$ represents the nirmatrelvir source function in the body, corresponding to a twice daily administration regimen (the time unit being $\tau = 1$ minute):

\begin{linenomath}
\begin{equation*}
  S(t,r) =
  \begin{cases}
  	K(r), & \text{if $\text{mod}(t, 12 \cdot 60) = 0$} \\
  	0, & \text{otherwise.}
  \end{cases}
\end{equation*}
\end{linenomath}

The precise value of $K(r)$ is discussed in the following section dedicated to parameter configuration.
    
Naturally, the above set of equations holds primarily for \textit{in vivo} scenarios. In case of \textit{in vitro} experiments one might, for example, consider a simpler, constant presence of nirmatrelvir.

\subsection{Parametrization}

The configuration of the stochastic ABM state space and the PDE layer describing SARS--CoV--2 infection had been given in \cite{hybrid-PDE-ABM-1}. Here we set the parameter values that are related to the (new) calibrated layer representing Paxlovid-based antiviral therapy.

\begin{itemize}

    \item \textbf{Drug removal rates: $\mu_c(r), \mu_N(r)$.} The main principle here is to find a configuration that guarantees best fit to clinical data. As we do not have direct information on the $\mu_c, \mu_N$ coefficients, we deduce them indirectly by using frequently measured nirmatrelvir blood concentration values communicated in \cite{nirmatrelvir-decay-data}. Of course, this argument raises the question whether it is reasonable to use blood concentration values to estimate local drug concentrations in the lung -- the validity of this approach is reassured by the results of \cite{invivo-drugconcentration-est}. The $\mu_c(r)$ and $\mu_N(r)$ coefficients were set using \textit{Mathematica}. We note that in the process we also benefited from a priori information on ritonavir from \cite{nirmatrelvir-decay-data} and \cite{ritonavir-fent}: we used that in non-ritonavir-boosted cases the active component is apparently metabolised $3$--$4$ times faster. The \textit{Mathematica} notebook is available in our public Github repository \cite{zenodo-github-project}.
    
    \item \textbf{Efficacy: $\eta_N$.} Applying a rather classical approach, in our model $\eta_N$ is defined by means of a Hill function -- the parameters of the latter are set precisely to obtain an efficacy of $50\%$ when the drug concentration takes the value of EC$_{50}$ for nirmatrelvir w.r.t. SARS--CoV--2 (the latter parameter is approximately $62$ nM according to \cite{fda-fact-sheet}). Formally, $\eta_N$ is defined as
    
    \begin{linenomath}
    \begin{equation}
    \label{eta-hill}
        \eta_N (N(t)) = \frac{1}{1 + \frac{\text{EC}_{50}}{N(t)}} .
    \end{equation}
    \end{linenomath}
    
    For simplicity, we use the notation $N(t)$ for drug concentration whether it is understood in nanomolars or in nanogramms per millilitre. Our implementation internally takes care of conversions when necessary due to data arriving from different sources.
    
\end{itemize}

The most important parameter values are summarized in Table \ref{drugparameters}.

\begin{table}[H]
\caption{Parameter configuration is primarily based upon best fit to actual data communicated in \cite{nirmatrelvir-decay-data}. Previously existing parameters are defined in \cite{hybrid-PDE-ABM-1}.\label{drugparameters}}
\begin{tabularx}{\textwidth}{CCCCC}
\toprule
\textbf{Symbol} & \textbf{Parameter} & \textbf{Unit} & \textbf{Ritonavir--boosted} & \textbf{Value} \\
\midrule
$\mu_c(r)$ & drug removal rate & $\tau^{-1}$ & false & 0.015 \\
 & in the stomach & & true & 0.005 \\
\\
$\mu_N(r)$ & drug removal rate & $\tau^{-1}$ & false & 0.013 \\
 & in the lung & & true & 0.004 \\
\\
$K(r)$ & drug source & $\text{ng}/\text{ml}/\tau$ & false & 1800 \\
 & in the stomach & & true & 6800 \\
\bottomrule
$\tau$: time unit
\end{tabularx}

\end{table}

\subsection{Implementation}

\begin{figure}
	\centering
	\includegraphics[scale=0.18]{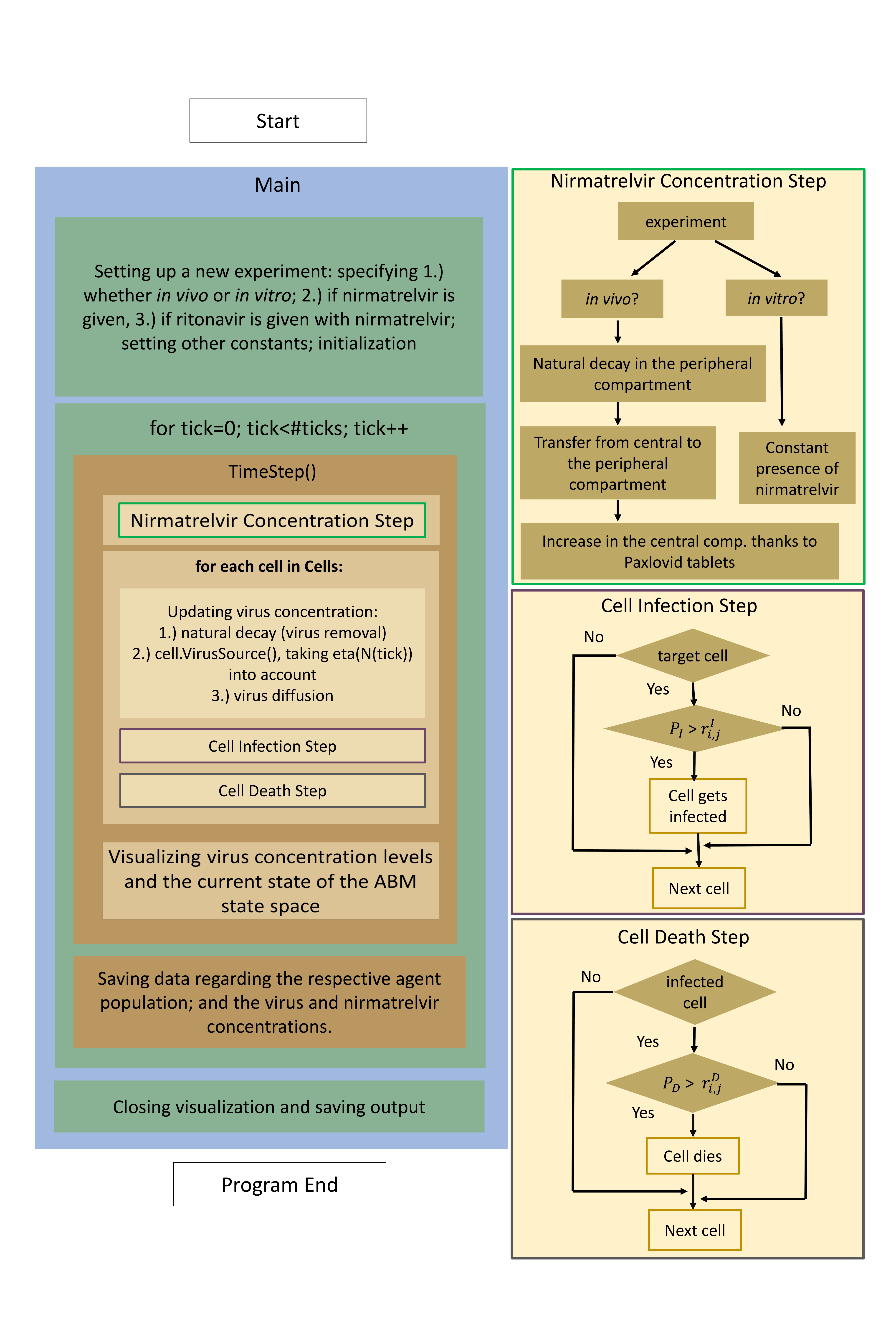}
	\caption[Algorithm]{The program flow diagram of the PDE-ABM model's implementation based on HAL {\cite{Bravo2020}}.}
	\label{fig:flowchart}
\end{figure}

The present work is a direct continuation of \cite{hybrid-PDE-ABM-1} and hence its technical foundations and principles remain unchanged. For the sake of compactness we avoid repetitive details -- here we limit ourselves to summarizing the extended structure of our updated software in a flowchart, see Figure~\ref{fig:flowchart}.

Our numerical simulations are based on a free and open source java software package, HAL (Hybrid Automata Library) \cite{Bravo2020}; our source code is publicly accessible in the Github repository \cite{zenodo-github-project}.

\section{Results}

\subsection{Replication of \textbf{in vitro} pharmacometrics of Paxlovid}

The initial step in identifying and testing clinically promising antiviral drugs consists in performing a great number of \textit{in vitro} experiments evaluating their overall effects. We begin with this straightforward approach, too. In this first scenario we simulate a series of experiments corresponding to \textit{in vitro} cases with different nirmatrelvir concentrations -- all these configurations are otherwise identical in every other aspect. We simulate the course of SARS--CoV--2 infection over the course of four days and we compare our computer-generated predictions with real-life observations obtained by scientific experiments assessing nirmatrelvir. Specifically, we consider Figure 3D in \cite{Mpro-inhibitor} -- here the authors evaluate PF-07321332 inhibition for (among other viruses) SARS--CoV--2 in viral-induced CPE assays, and their results are given for a series of different drug concentration values.

\begin{figure}[H]
    \centering
    \includegraphics[width=0.6\textwidth]{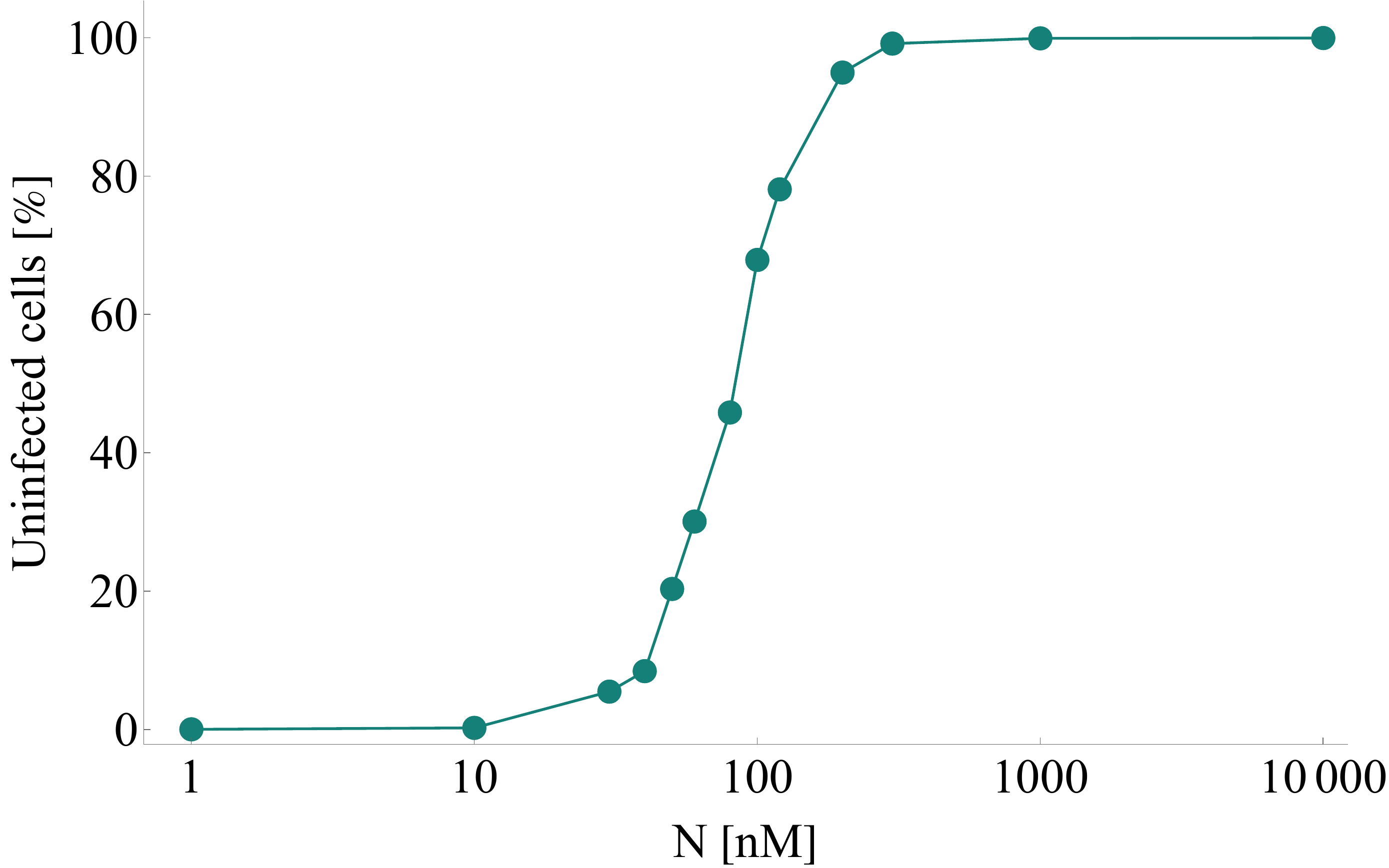}
    \caption{A simulated series of \textit{in vitro} experiments with increasing initial nirmatrelvir concentrations. Concentration levels are assumed to be constant throughout the entire course of each experiment. Every simulation follows the emerging infection dynamics for $4$ days. SARS--CoV--2 infection and nirmatrelvir treatment are initialized simultaneously. Our computer-generated predictions correspond reassuringly to real-life scientific measurements assessing infection inhibition of PF-07321332, see Figure 3D in \cite{Mpro-inhibitor}.}
    \label{invitrohealthycurve}
\end{figure}
    
Figure~\ref{invitrohealthycurve} demonstrates a notable resemblance to Figure 3D in \cite{Mpro-inhibitor}. Key features of inhibition efficacy match in a reassuring way: the characteristic shape itself of the calculated curve looks identical to its clinical counterpart, and the numbers connected to the main concentration window (approximately between $10$ nM and $300$ nM) corresponding to tangible increase are essentially the same, too.

While there is a clear match between clinical data and our calculated results, we highlight that there is a natural limit to accuracy due to simple lack of data. Both simulated and real-life outcomes naturally depend on key features such as the number of days the experiment went on for or the complete resolution of the state space (i.e. the total number of cells). While the supplementary material of \cite{Mpro-inhibitor} suggests that the authors mostly considered time intervals corresponding to $3$--$5$ days, several parameter values are either unknown by nature or have not been disclosed.

\subsection{Exploring \textbf{in vivo} pharmacometrics of Paxlovid}

In this section we present and explain our most significant computational results representing simplified \textit{in vivo} cases. We explore a series of scenarios with various configurations, one basic feature remains unchanged however in all of them: we assume that the most fundamental instructions given in Paxlovid's documentation \cite{fda-fact-sheet} are followed at least for its nirmatrelvir component. Technically this means that once the patient starts taking Paxlovid, they steadily take at least nirmatrelvir for $5$ days straight, $1$ dose every $12$ hours. Since Paxlovid is essentially nirmatrelvir co-packaged with ritonavir, technically it is possible that a patient -- either consciously because of an existing drug allergy or simply because of forgetfulness -- takes only nirmatrelvir, without the added benefits of ritonavir. This degree of freedom is allowed and investigated throughout the simulations. Some other combinations and scenarios were excluded due to lack of data, again others were omitted simply because of the limited scope of the article.

\begin{Remark} \normalfont
The inherent, rather sharp distinction between the \textit{in vitro} and \textit{in vivo} clinical categories becomes notably smoother in the simulated context of our mathematical model. Figuratively speaking, we perform \textit{in vivo} experiments "as if they were" \textit{in vitro} in the sense that we have full control over (and full information on) which specific biological or anatomical processes are included and which ones are left out. Our framework is called hybrid because of the different mathematical theories it unites, but it proves to be hybrid in this point of view as well.
\end{Remark}

\subsubsection{In silico testing of immediate Paxlovid-based intervention}

\begin{figure}[H]
\centering
\begin{minipage}{0.89\textwidth}
    \captionsetup[subfigure]{justification=centering}
    \centering
    \begin{subfigure}[b]{0.43\textwidth}
        \includegraphics[width=\textwidth]{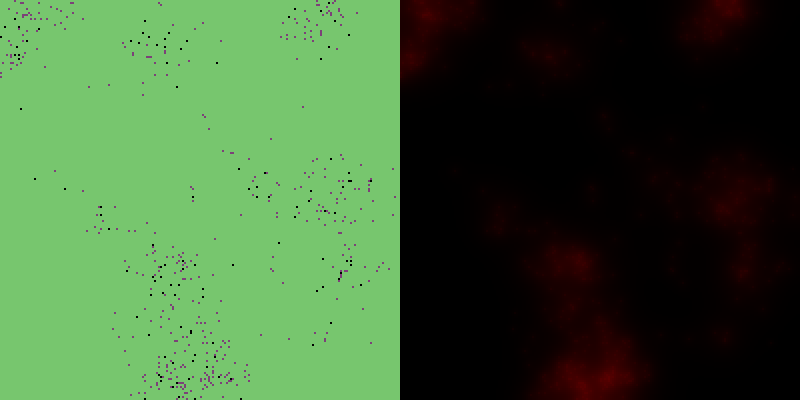} 
        \label{no-drug-day1}
        \vspace{-\baselineskip}
        \caption{}
    \end{subfigure}
     \begin{subfigure}[b]{0.43\textwidth}
        \includegraphics[width=\textwidth]{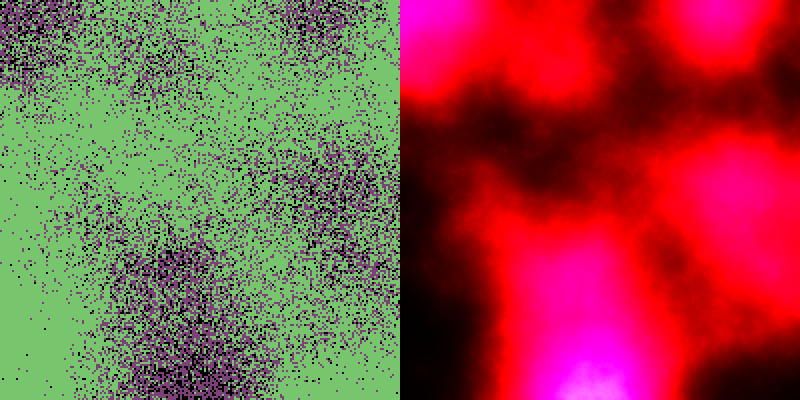}
        \label{no-drug-day2} 
        \vspace{-\baselineskip}
        \caption{}
    \end{subfigure}
    \par\bigskip
    \begin{subfigure}[b]{0.43\textwidth}
        \includegraphics[width=\textwidth]{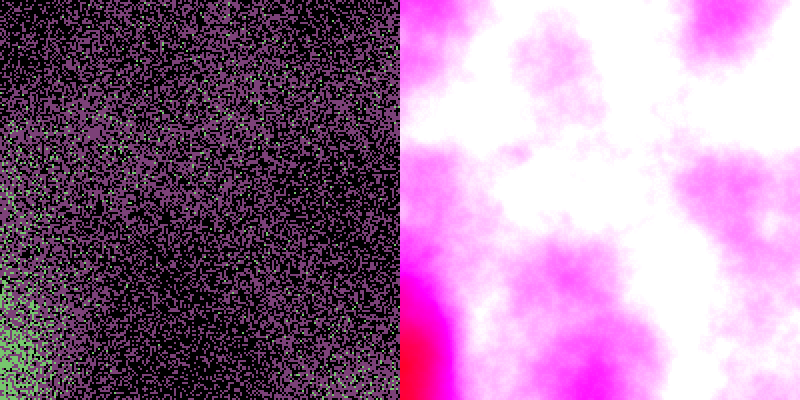} 
        \label{no-drug-day3}
        \vspace{-\baselineskip}
        \caption{}
    \end{subfigure}
     \begin{subfigure}[b]{0.43\textwidth}
        \includegraphics[width=\textwidth]{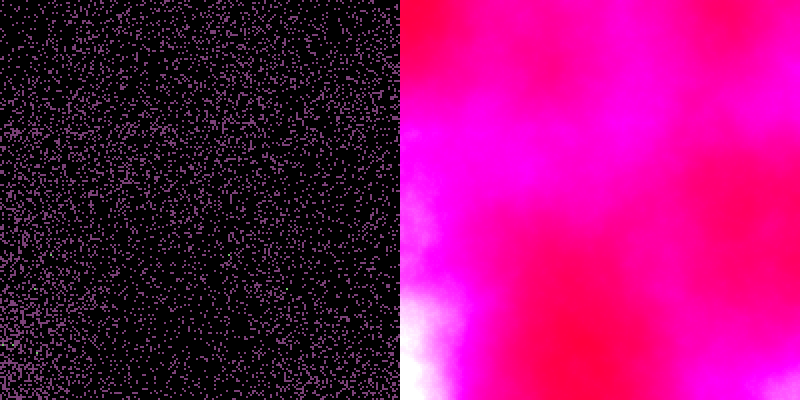}
        \label{no-drug-day4} 
        \vspace{-\baselineskip}
        \caption{}
    \end{subfigure}
\end{minipage}
\hspace{0.005\textwidth}
\begin{minipage}{0.09\textwidth}
    \includegraphics[width=0.7\textwidth]{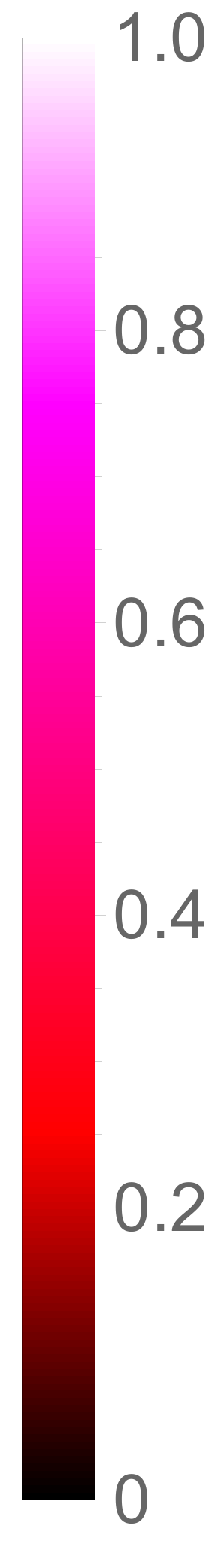}
    \vspace{0.3cm}
\end{minipage}
    \caption{Simulated spatiotemporal solutions captured (\textbf{a}) 24 hours, (\textbf{b}) 48 hours, (\textbf{c}) 72 hours, and (\textbf{d}) 96 hours after SARS--CoV--2 infection. \textbf{No antiviral intervention} took place in this case. The cellular state spaces are depicted on the left in all four subfigures; uninfected, infected and dead cells are denoted by green, purple, and black squares, respectively. Virus concentration values are shown on the right. The colour bar is understood in virions per unit space.}
    \label{no-drug-spatial}
\end{figure}

\begin{figure}[H]
\centering
\begin{minipage}{0.89\textwidth}
    \captionsetup[subfigure]{justification=centering}
    \centering
    \begin{subfigure}[b]{0.43\textwidth}
        \includegraphics[width=\textwidth]{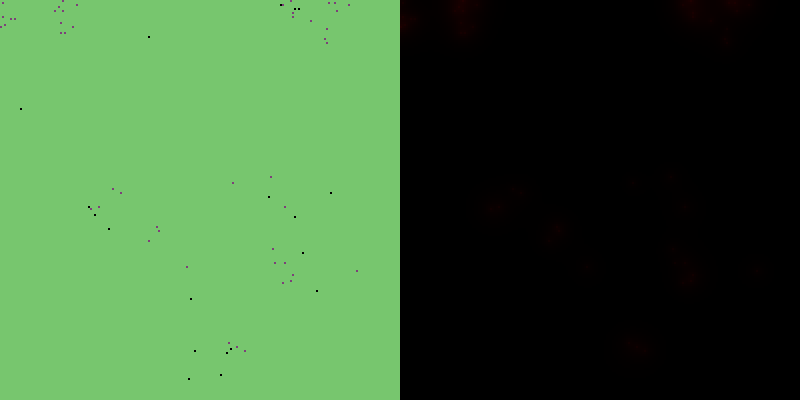} 
        \label{nirm-only-day1}
        \vspace{-\baselineskip}
        \caption{}
    \end{subfigure}
     \begin{subfigure}[b]{0.43\textwidth}
        \includegraphics[width=\textwidth]{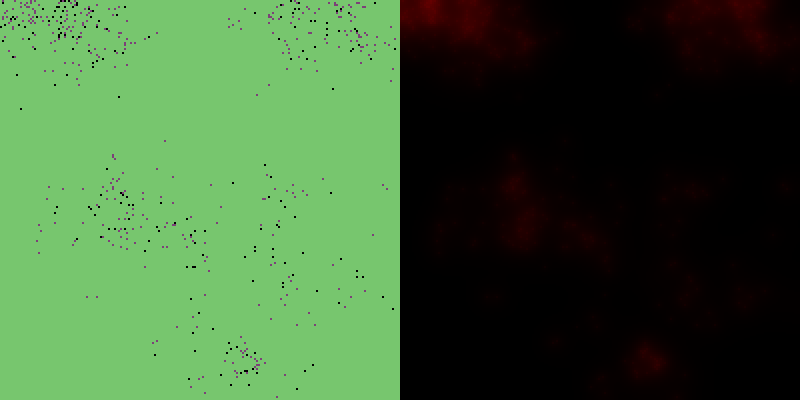} 
        \label{nirm-only-day2} 
        \vspace{-\baselineskip}
        \caption{}
    \end{subfigure}
    \par\bigskip
    \begin{subfigure}[b]{0.43\textwidth}
        \includegraphics[width=\textwidth]{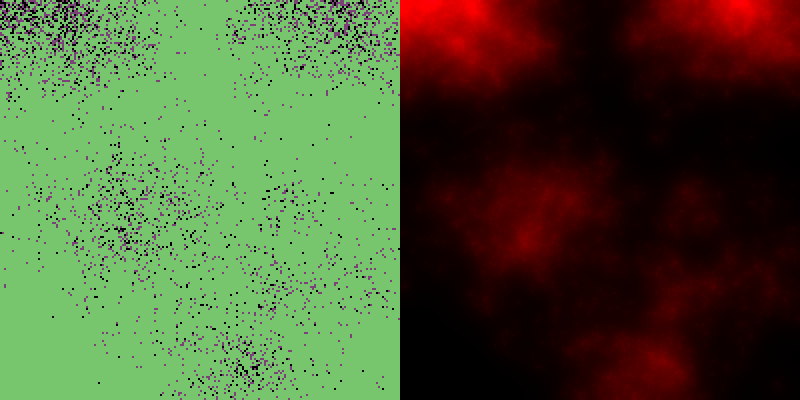} 
        \label{nirm-only-day3}
        \vspace{-\baselineskip}
        \caption{}
    \end{subfigure}
     \begin{subfigure}[b]{0.43\textwidth}
        \includegraphics[width=\textwidth]{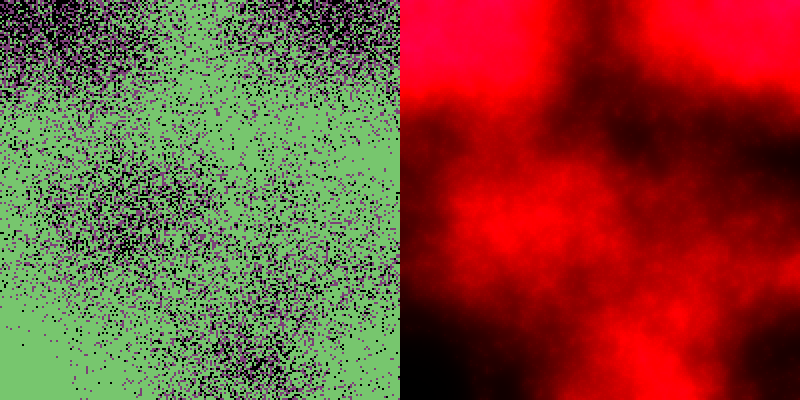} 
        \label{nirm-only-day4}
        \vspace{-\baselineskip}
        \caption{}
    \end{subfigure}
\end{minipage}
\hspace{0.005\textwidth}
\begin{minipage}{0.09\textwidth}
    \phantom{\includegraphics[width=0.7\textwidth]{images/colorbar_font26_wide.pdf}}
    \vspace{0.3cm}
\end{minipage}
    \caption{Simulated spatiotemporal solutions captured (\textbf{a}) 24 hours, (\textbf{b}) 48 hours, (\textbf{c}) 72 hours, and (\textbf{d}) 96 hours after SARS--CoV--2 infection and simultaneous treatment with nirmatrelvir. In this case \textbf{nirmatrelvir was given without ritonavir}, intervention took place with no delay. The cellular state spaces are depicted on the left in all four subfigures; uninfected, infected and dead cells are denoted by green, purple, and black squares, respectively. Virus concentration values are shown on the right according to the scale in Figure~\ref{no-drug-spatial}.}
    \label{nirm-only-spatial}
\end{figure}

\begin{figure}[H]
\centering
\begin{minipage}{0.89\textwidth}
    \captionsetup[subfigure]{justification=centering}
    \centering
    \begin{subfigure}[b]{0.43\textwidth}
        \includegraphics[width=\textwidth]{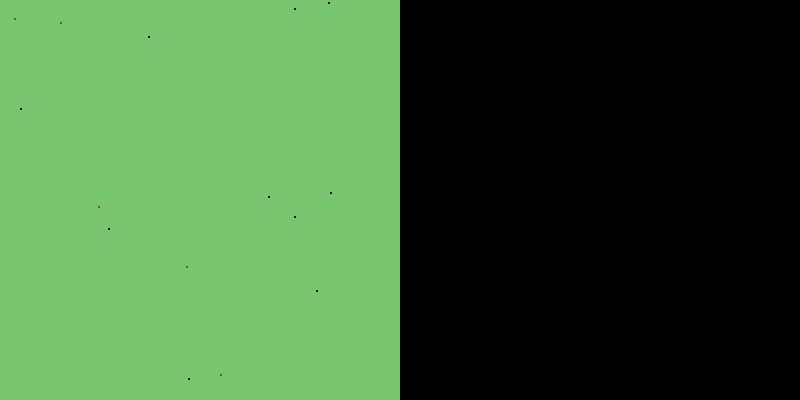} 
        \label{rito-nirm-day1}
        \vspace{-\baselineskip}
        \caption{}
    \end{subfigure}
     \begin{subfigure}[b]{0.43\textwidth}
        \includegraphics[width=\textwidth]{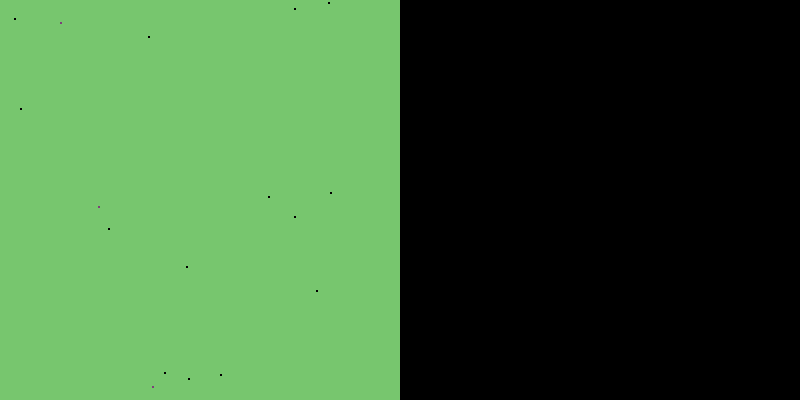} 
        \label{rito-nirm-day2} 
        \vspace{-\baselineskip}
        \caption{}
    \end{subfigure}
    \par\bigskip
    \begin{subfigure}[b]{0.43\textwidth}
        \includegraphics[width=\textwidth]{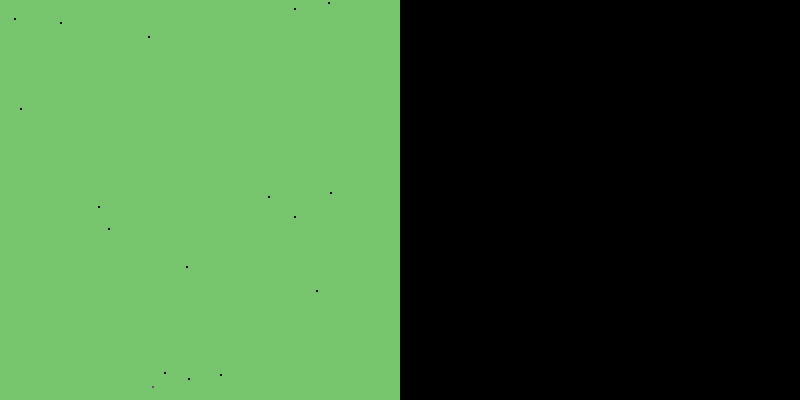} 
        \label{rito-nirm-day3}
        \vspace{-\baselineskip}
        \caption{}
    \end{subfigure}
     \begin{subfigure}[b]{0.43\textwidth}
        \includegraphics[width=\textwidth]{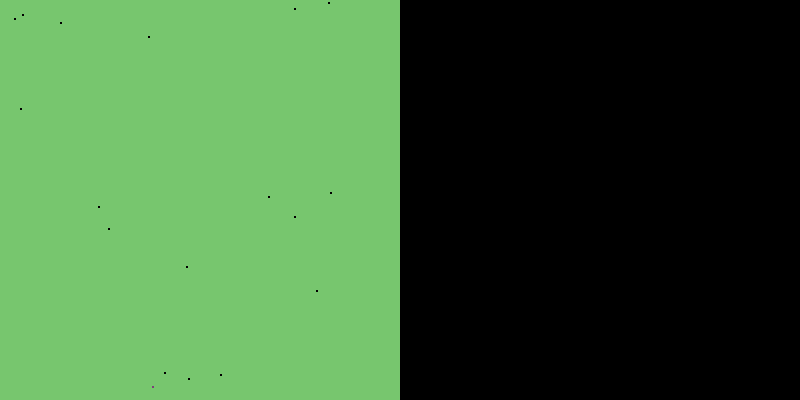} 
        \label{rito-nirm-day4}
        \vspace{-\baselineskip}
        \caption{}
    \end{subfigure}
\end{minipage}
\hspace{0.005\textwidth}
\begin{minipage}{0.09\textwidth}
    \phantom{\includegraphics[width=0.7\textwidth]{images/colorbar_font26_wide.pdf}}
    \vspace{0.3cm}
\end{minipage}
    \caption{Simulated spatiotemporal solutions captured (\textbf{a}) 24 hours, (\textbf{b}) 48 hours, (\textbf{c}) 72 hours, and (\textbf{d}) 96 hours after SARS--CoV--2 infection and simultaneous treatment with Paxlovid. In this case \textbf{ritonavir-boosted nirmatrelvir was given}, i.e. official instructions regarding Paxlovid were followed. Intervention took place with no delay. The cellular state spaces are depicted on the left in all four subfigures; uninfected, infected and dead cells are denoted by green, purple, and black squares, respectively. Virus concentration values are shown on the right according to the scale in Figure~\ref{no-drug-spatial}.}
    \label{rito-nirm-spatial}
\end{figure}

We begin by simulating three basic scenarios and observing the respective outcomes. Figure~\ref{no-drug-spatial} shows the course of SARS--CoV--2 infection assuming no antiviral intervention, Figure~\ref{nirm-only-spatial} follows a case where the patient takes nirmatrelvir only (i.e. the main acting component of Paxlovid, without the benefits of ritonavir), while Figure~\ref{rito-nirm-spatial} represents the scenario where Paxlovid is taken exactly according to official instructions.

\begin{figure}[H]
    \captionsetup[subfigure]{justification=centering}
    \centering
    \begin{subfigure}[b]{0.48\textwidth}
        \includegraphics[width=\textwidth]{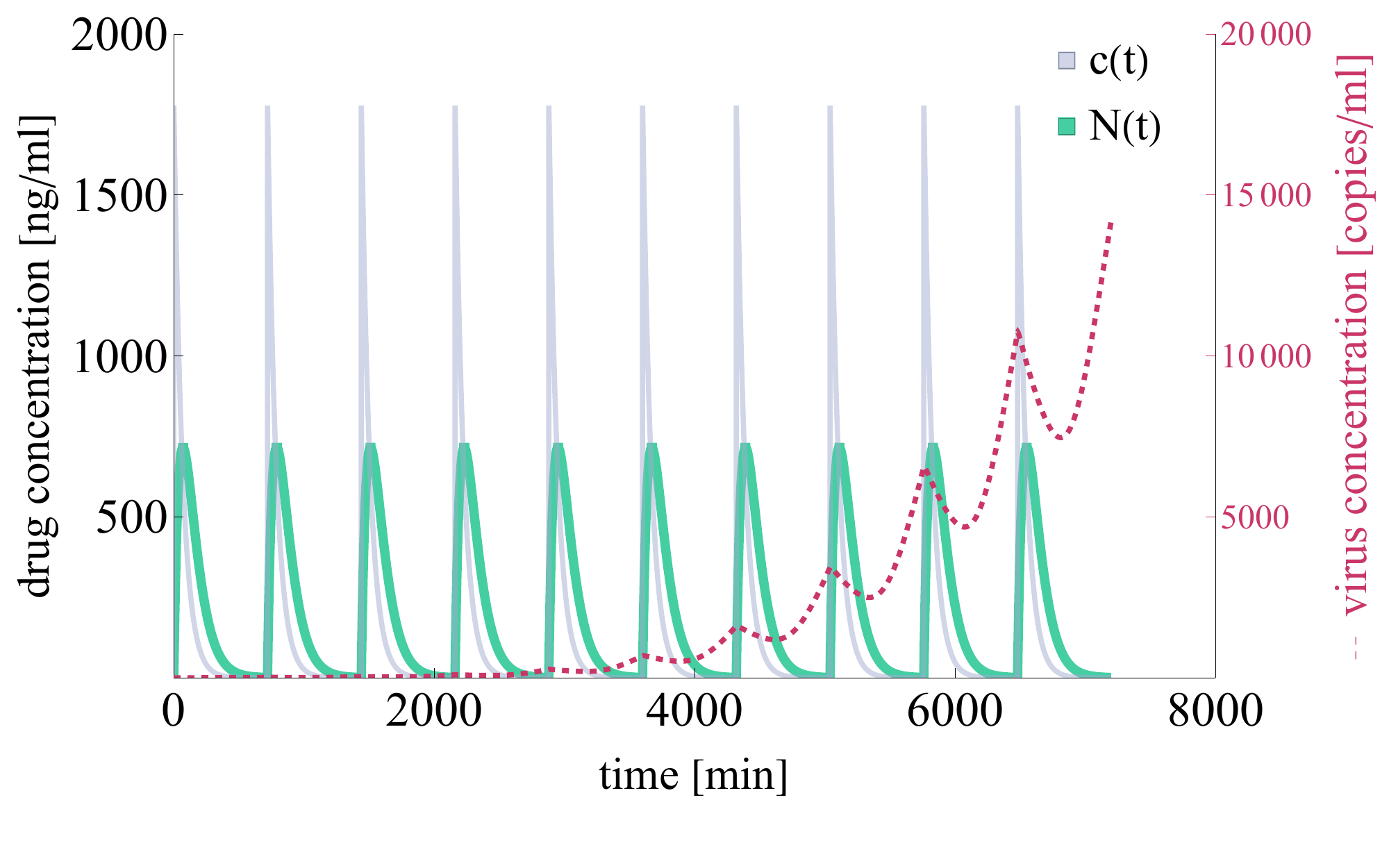} 
        \label{nirmatrelvir-only}
        \vspace{-2\baselineskip}
        \caption{nirmatrelvir only}
    \end{subfigure}
     \begin{subfigure}[b]{0.48\textwidth}
        \includegraphics[width=\textwidth]{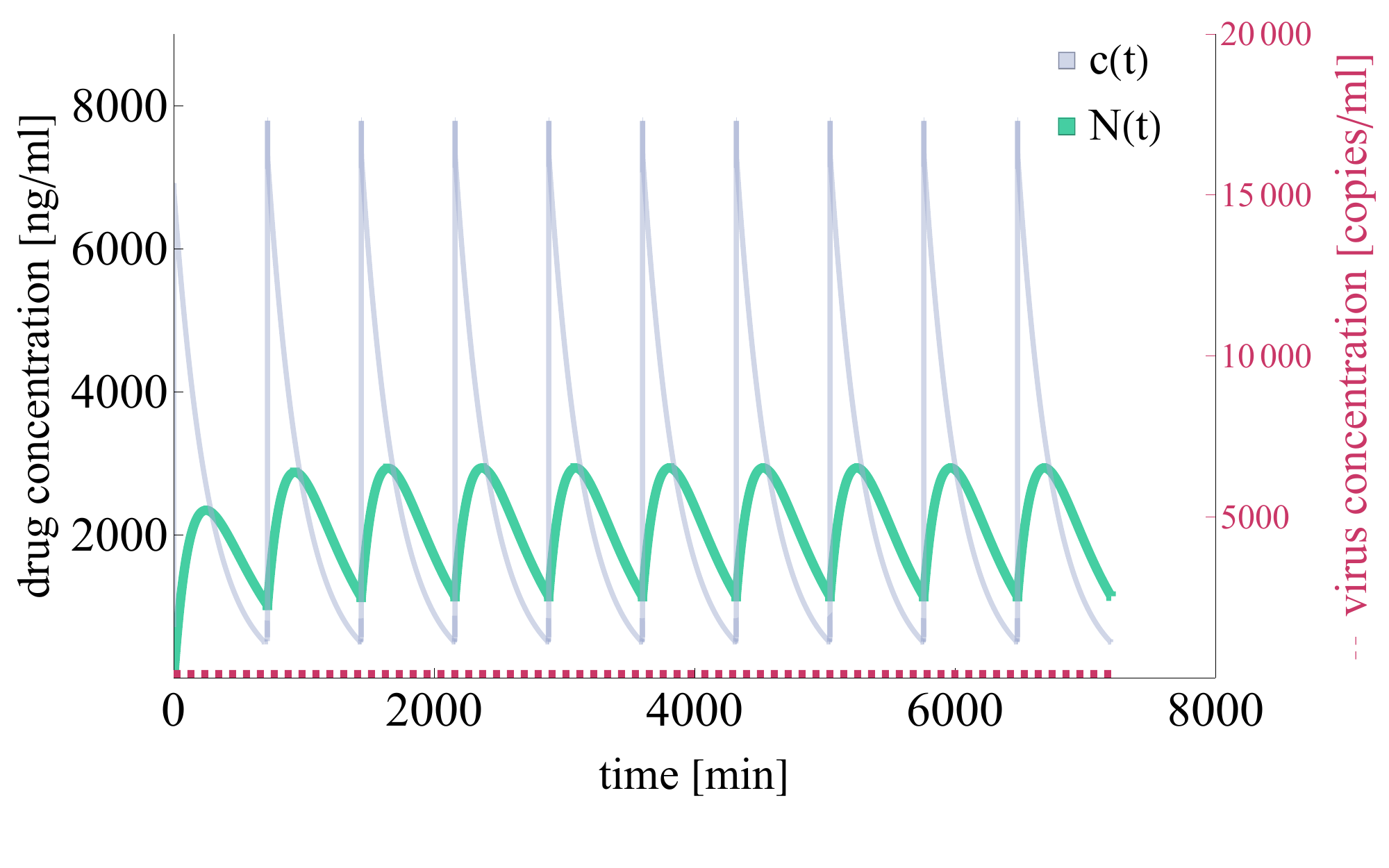}
        \label{ritonavir-boosted} 
        \vspace{-2\baselineskip}
        \caption{Paxlovid}
    \end{subfigure}
    \caption{Integrated virus concentration and nirmatrelvir concentration levels for two different scenarios representing nirmatrelvir-based intervention. Subfigure (\textbf{a}) shows the simulated outcome of applying nirmatrelvir without ritonavir, while subfigure (\textbf{b}) depicts the results of rigorous treatment with Paxlovid (ritonavir-boosted nirmatrelvir). SARS--CoV--2 virus concentrations are coloured in red (shown dashed), nirmatrelvir concentration levels -- $N(t)$ and $c(t)$ -- are depicted in sea green and light purple, respectively.}
    \label{n-o-vs-rbn}
\end{figure}

For the latter two cases we plot total virus concentration and nirmatrelvir concentrations both at the first level of metabolism in the body and locally at the epithelial lung cells, the results are shown in Figure~\ref{n-o-vs-rbn}.

In Figure~\ref{n-o-vs-rbn}, both the integrated virus and drug concentration values are noteworthy. Firstly, we highlight that nirmatrelvir concentration levels clearly correspond to Figure 2A in \cite{nirmatrelvir-decay-data} -- this means that our simulations (both in the ritonavir-boosted and in the nirmatrelvir-only case) are running with highly realistic nirmatrelvir concentration levels. Secondly, our computational results correspond to straightforward, basic expectations suggested by the packaging of Paxlovid. In more detail; on the one hand nirmatrelvir in itself seems to be insufficient to control the infection (which explains why Paxlovid does not simply consist of nirmatrelvir tablets), and on the other hand, ritonavir-boosted nirmatrelvir is apparently capable to stop infection entirely (which is in accordance with the simple fact that Paxlovid is an authorized drug of great promise).

\subsubsection{Evaluating the effects of treatment delay}

The previous section's premise was similar to a classical \textit{in vitro} configuration -- in this original default case, infection and treatment began simultaneously. In order to make our model more realistic, here we introduce and explore a new degree of freedom: treatment delay.

We begin by exploring how simulated predictions seen in Figures~\ref{nirm-only-spatial}, \ref{rito-nirm-spatial}, and  \ref{n-o-vs-rbn} would change if Paxlovid tablets were given with a delay.

\begin{figure}[H]
\centering
\begin{minipage}{0.89\textwidth}
    \captionsetup[subfigure]{justification=centering}
    \centering
    \begin{subfigure}[b]{0.43\textwidth}
        \includegraphics[width=\textwidth]{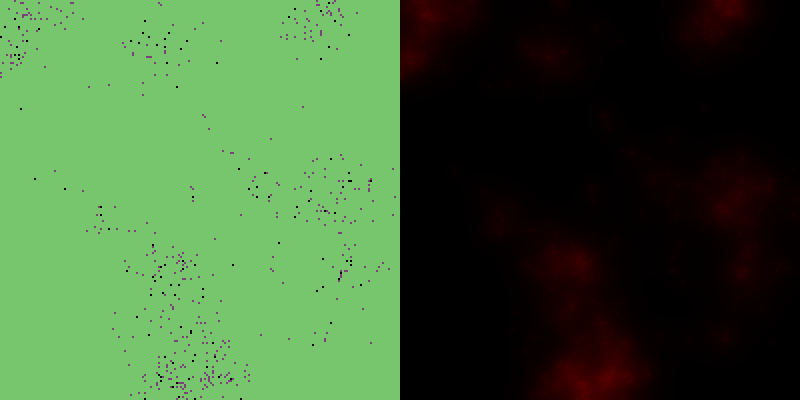} 
        \caption{}
    \end{subfigure}
     \begin{subfigure}[b]{0.43\textwidth}
        \includegraphics[width=\textwidth]{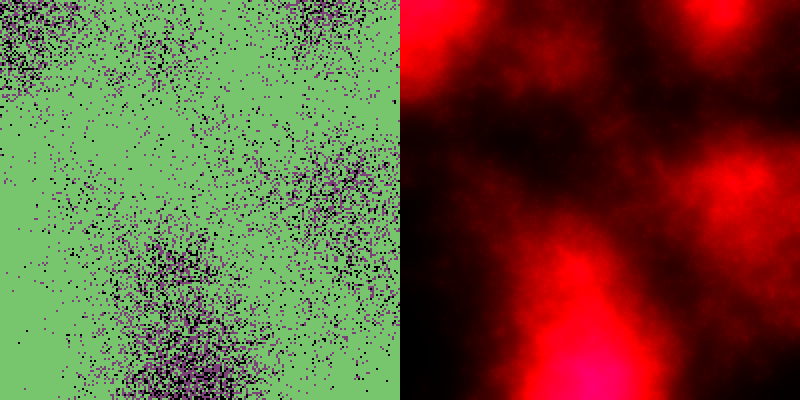}  
        \caption{}
    \end{subfigure}
    \par\bigskip
    \begin{subfigure}[b]{0.43\textwidth}
        \includegraphics[width=\textwidth]{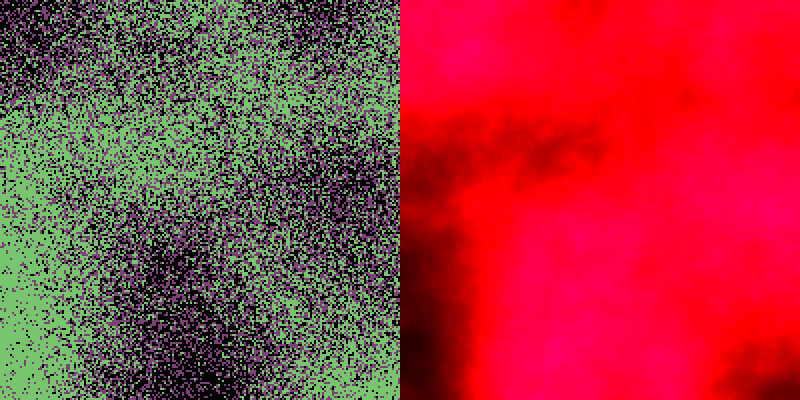}  
        \caption{}
    \end{subfigure}
     \begin{subfigure}[b]{0.43\textwidth}
        \includegraphics[width=\textwidth]{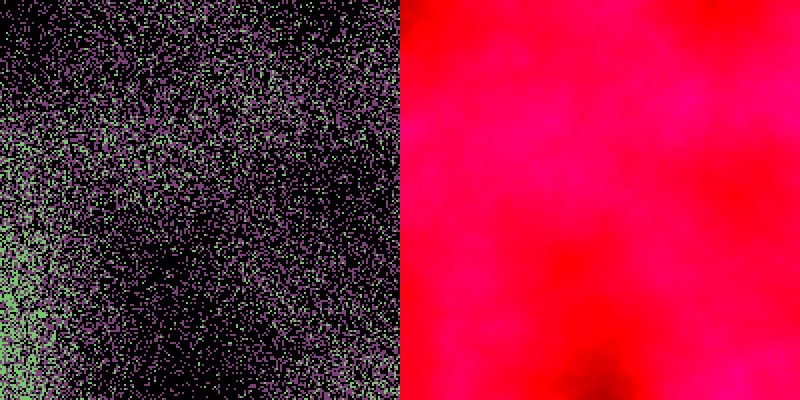}  
        \caption{}
    \end{subfigure}
    \par\bigskip
    \begin{subfigure}[b]{0.43\textwidth}
        \includegraphics[width=\textwidth]{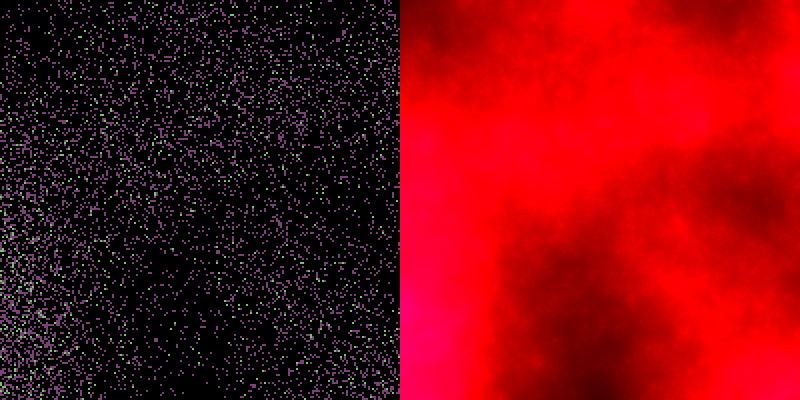}  
        \caption{}
    \end{subfigure}
     \begin{subfigure}[b]{0.43\textwidth}
        \includegraphics[width=\textwidth]{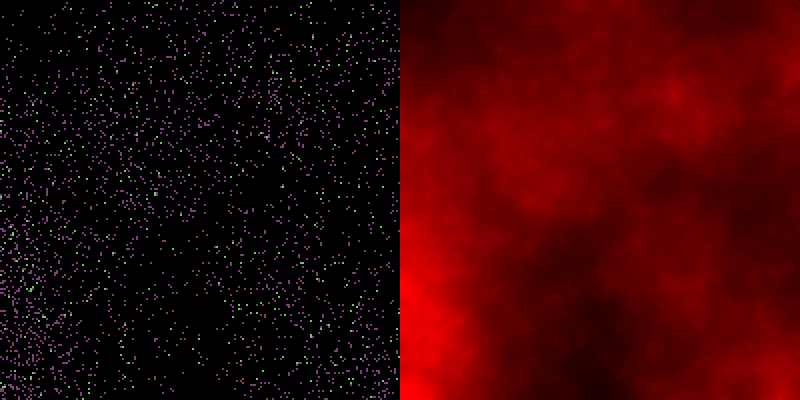}  
        \caption{}
    \end{subfigure}
\end{minipage}
\hspace{0.005\textwidth}
\begin{minipage}{0.09\textwidth}
    \phantom{\includegraphics[width=0.7\textwidth]{images/colorbar_font26_wide.pdf}}
    \vspace{0.3cm}
\end{minipage}
    \caption{Simulated spatiotemporal solutions captured (\textbf{a}) 24 hours, (\textbf{b}) 48 hours, (\textbf{c}) 72 hours, (\textbf{d}) 96, (\textbf{e}) 120, and (\textbf{f}) 144 hours after SARS--CoV--2 infection and delayed treatment with nirmatrelvir. In this case \textbf{nirmatrelvir was given without ritonavir}, intervention took place after a 36-hour delay. The cellular state spaces are depicted on the left in all four subfigures; uninfected, infected and dead cells are denoted by green, purple, and black squares, respectively. Virus concentration values are shown on the right according to the scale in Figure~\ref{no-drug-spatial}.}
    \label{nirm-only-spatial-36hdelay}
\end{figure}

\begin{figure}[H]
\centering
\begin{minipage}{0.89\textwidth}
    \captionsetup[subfigure]{justification=centering}
    \centering
    \begin{subfigure}[b]{0.43\textwidth}
        \includegraphics[width=\textwidth]{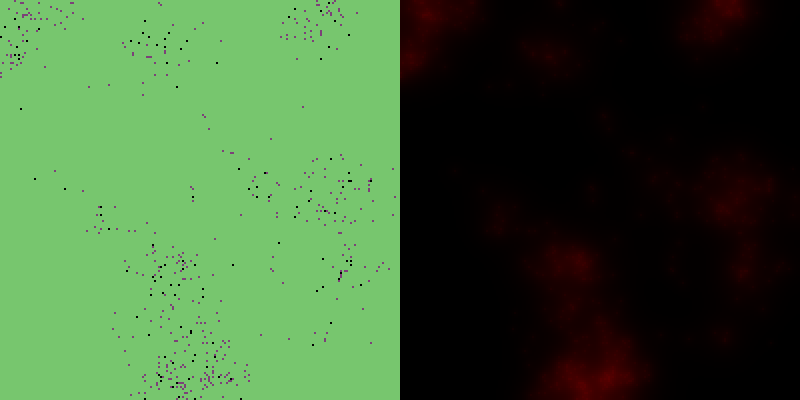} 
        \caption{}
    \end{subfigure}
     \begin{subfigure}[b]{0.43\textwidth}
        \includegraphics[width=\textwidth]{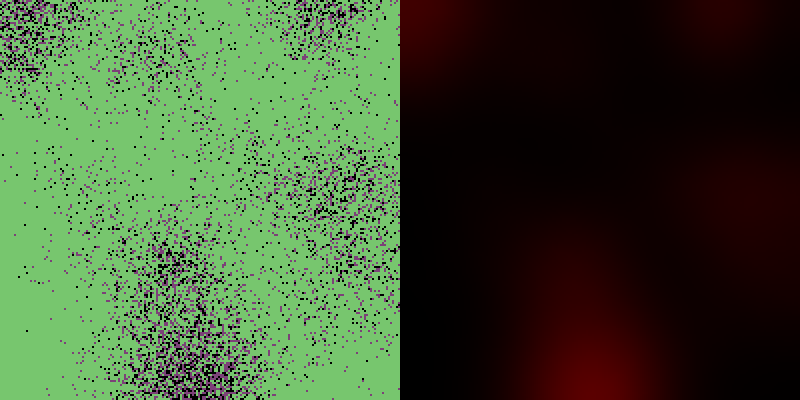}  
        \caption{}
    \end{subfigure}
    \par\bigskip
    \begin{subfigure}[b]{0.43\textwidth}
        \includegraphics[width=\textwidth]{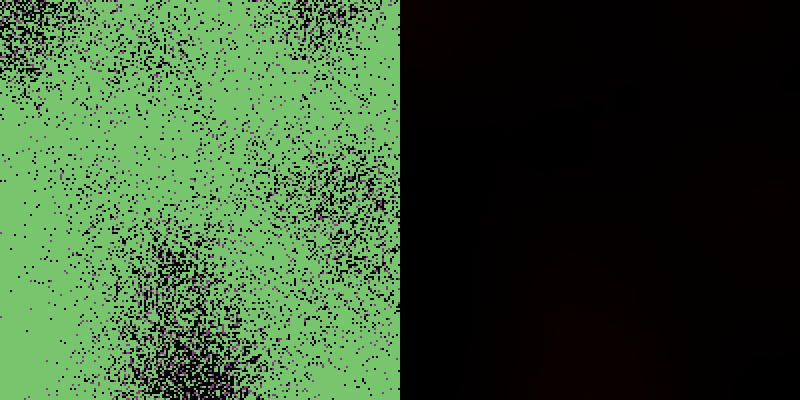}  
        \caption{}
    \end{subfigure}
     \begin{subfigure}[b]{0.43\textwidth}
        \includegraphics[width=\textwidth]{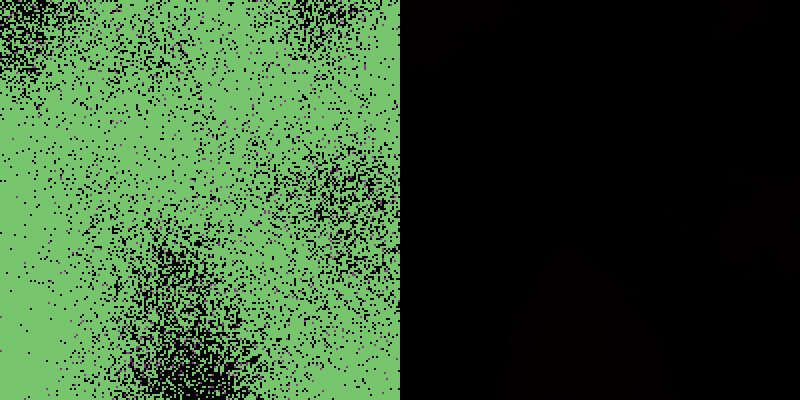}  
        \caption{}
    \end{subfigure}
    \par\bigskip
    \begin{subfigure}[b]{0.43\textwidth}
        \includegraphics[width=\textwidth]{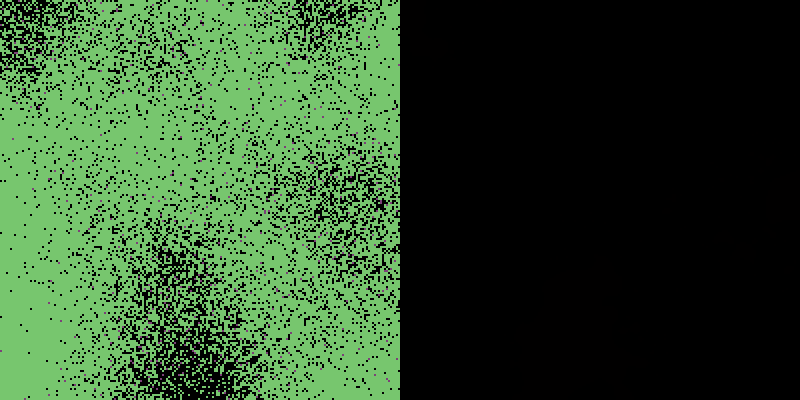}  
        \caption{}
    \end{subfigure}
     \begin{subfigure}[b]{0.43\textwidth}
        \includegraphics[width=\textwidth]{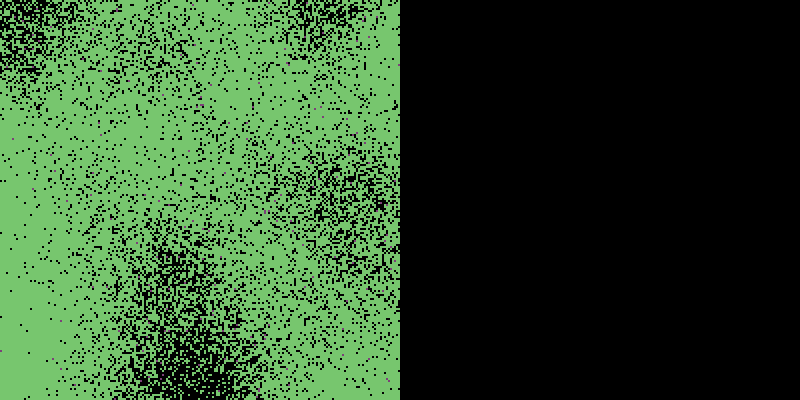}  
        \caption{}
    \end{subfigure}
\end{minipage}
\hspace{0.005\textwidth}
\begin{minipage}{0.09\textwidth}
    \phantom{\includegraphics[width=0.7\textwidth]{images/colorbar_font26_wide.pdf}}
    \vspace{0.3cm}
\end{minipage}
    \caption{Simulated spatiotemporal solutions captured (\textbf{a}) 24 hours, (\textbf{b}) 48 hours, (\textbf{c}) 72 hours, (\textbf{d}) 96, (\textbf{e}) 120, and (\textbf{f}) 144 hours after SARS--CoV--2 infection and delayed treatment with Paxlovid. In this case \textbf{ritonavir-boosted nirmatrelvir was given}, i.e. official instructions regarding Paxlovid were followed. Intervention took place after a 36-hour delay. The cellular state spaces are depicted on the left in all four subfigures; uninfected, infected and dead cells are denoted by green, purple, and black squares, respectively. Virus concentration values are shown on the right according to the scale in Figure~\ref{no-drug-spatial}.}
    \label{rito-nirm-spatial-36hdelay}
\end{figure}

In particular, Figure~\ref{nirm-only-spatial-36hdelay} and Figure~\ref{rito-nirm-spatial-36hdelay} illustrate virus dynamical processes that are otherwise identical to the scenarios of Figure~\ref{nirm-only-spatial} and Figure~\ref{rito-nirm-spatial},  respectively, except for a $36$-hour delay in initiating nirmatrelvir-based treatment (this also means that we follow these cases for an overall longer time period). The ritonavir-boosted scenario is particularly interesting. Though the first $36$ hours see uninhibited virus spread, Figure~\ref{rito-nirm-spatial-36hdelay} confirms that Paxlovid can control infection relatively well even in this particular, less favorable scenario: after the first $2$ days there are almost no new cell infections at all, the only detectable change between the last four subfigures is infected cells gradually turning dead.

\begin{figure}[H]
    \captionsetup[subfigure]{justification=centering}
    \centering
    \begin{subfigure}[b]{0.48\textwidth}
        \includegraphics[width=\textwidth]{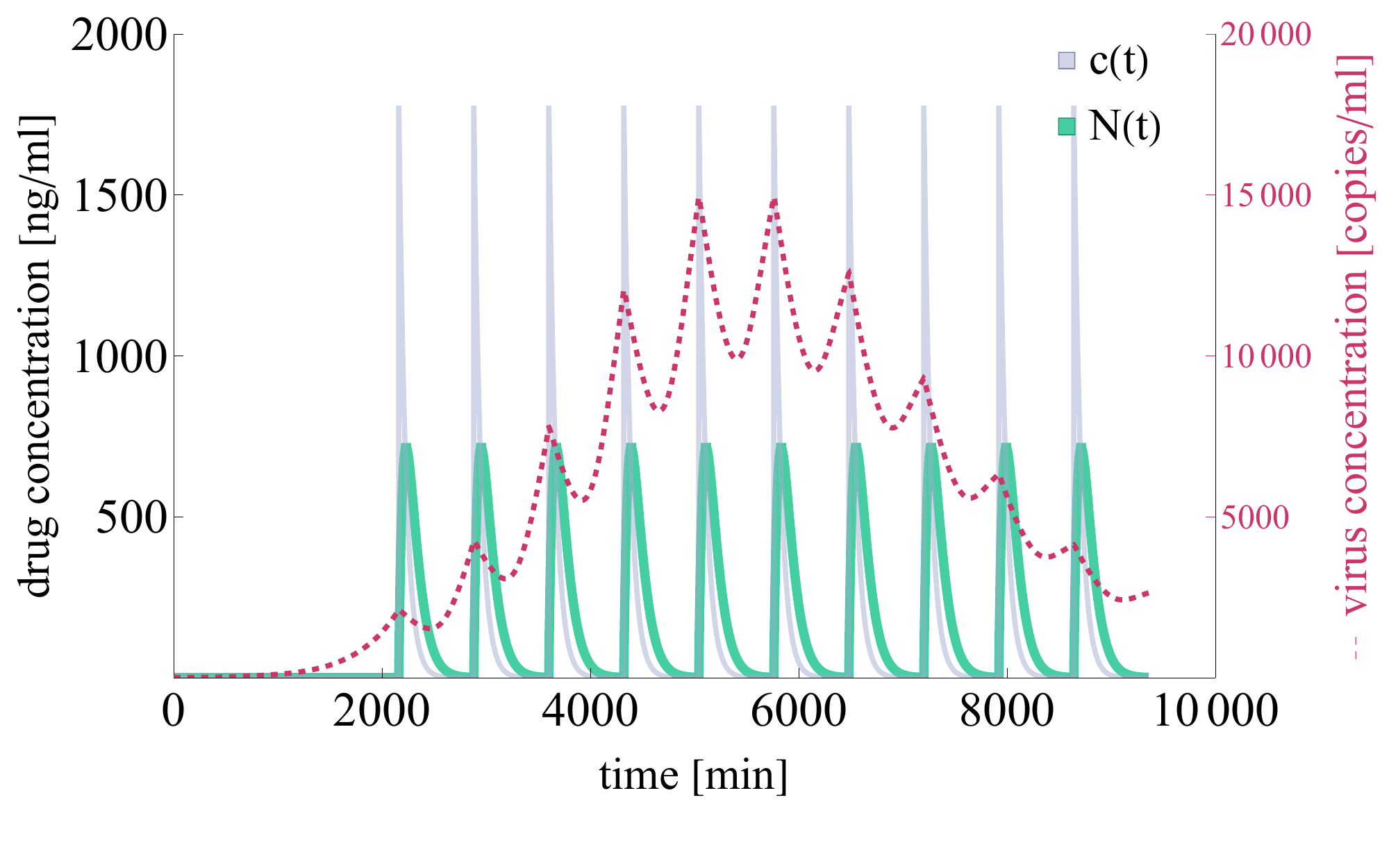} 
        \vspace{-2\baselineskip}
        \caption{nirmatrelvir only}
    \end{subfigure}
     \begin{subfigure}[b]{0.48\textwidth}
        \includegraphics[width=\textwidth]{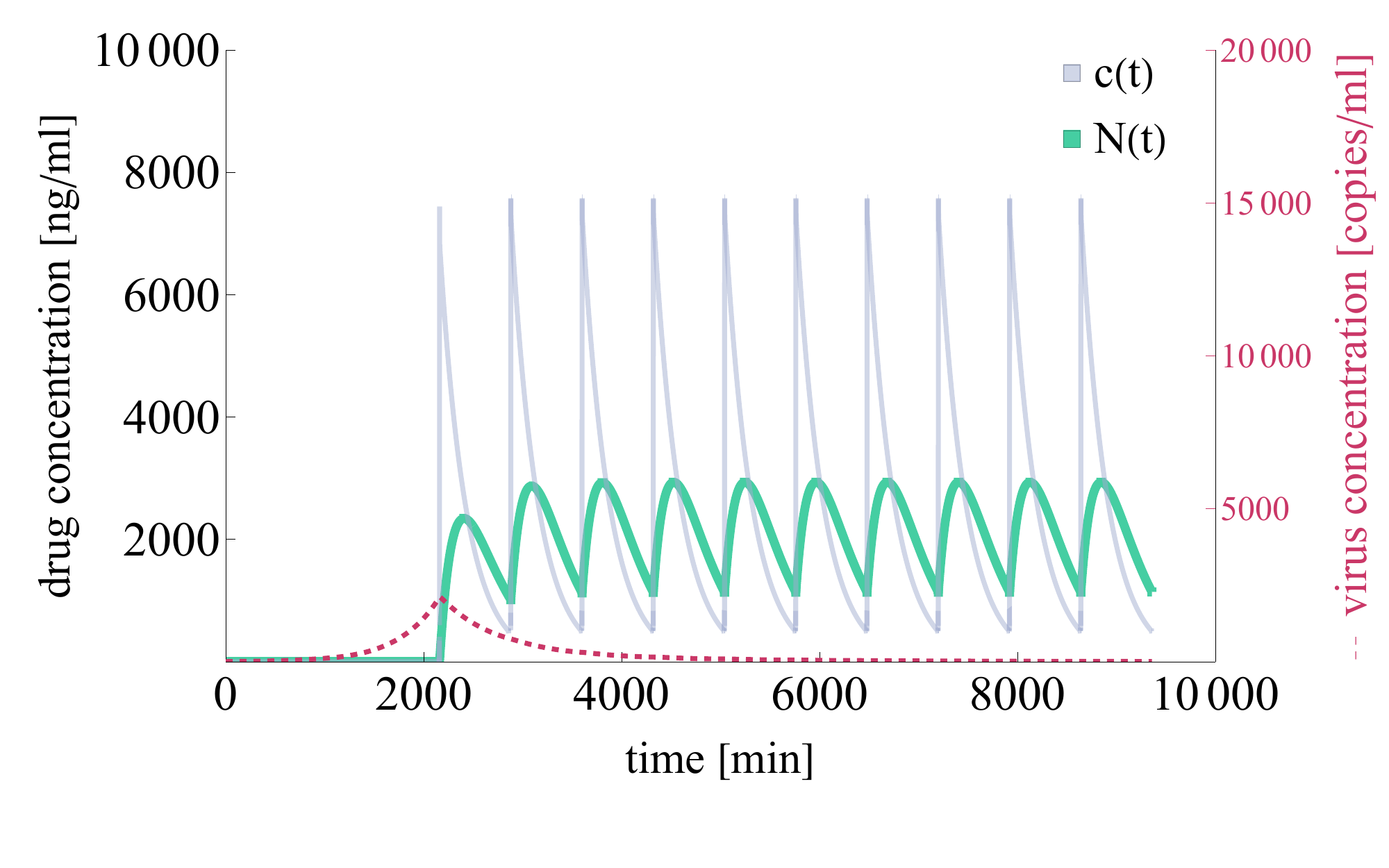}
        \vspace{-2\baselineskip}
        \caption{Paxlovid}
    \end{subfigure}
    \caption{Integrated virus concentration and nirmatrelvir concentration levels for two different scenarios representing nirmatrelvir-based intervention. In both cases tablets are given after a \textbf{36-hour delay} w.r.t infection initialization. Subfigure (\textbf{a}) shows the simulated outcome of applying nirmatrelvir without ritonavir, while subfigure (\textbf{b}) depicts the results of rigorous treatment with Paxlovid (ritonavir-boosted nirmatrelvir). SARS--CoV--2 virus concentrations are coloured in red (shown dashed), nirmatrelvir concentration levels -- $N(t)$ and $c(t)$ -- are depicted in sea green and light purple, respectively.}
    \label{n-o-vs-rbn-36hd}
\end{figure}

Similarly to Figure~\ref{n-o-vs-rbn}, Figure~\ref{n-o-vs-rbn-36hd} shows integrated virus concentration and nirmatrelvir concentration levels, but, naturally, considering a $36$-hour delay before Paxlovid is given.

Now we are ready to move on to this section's main purpose, namely, investigating outcomes and eventual averted tissue damage rates for a series of delay values with respect to the default (i.e. no delay) case. Note that with no particular immune response, total tissue damage (i.e. the ratio of cells that are either infected or already dead) after $5$ days reaches $100$ percent -- this means that the ratio of eventually remaining susceptible target cells after Paxlovid treatment corresponds precisely to the damage that is averted because of Paxlovid.

\begin{figure}[H]
    \captionsetup[subfigure]{justification=centering}
    \centering
    \begin{subfigure}[b]{0.48\textwidth}
        \includegraphics[width=\textwidth]{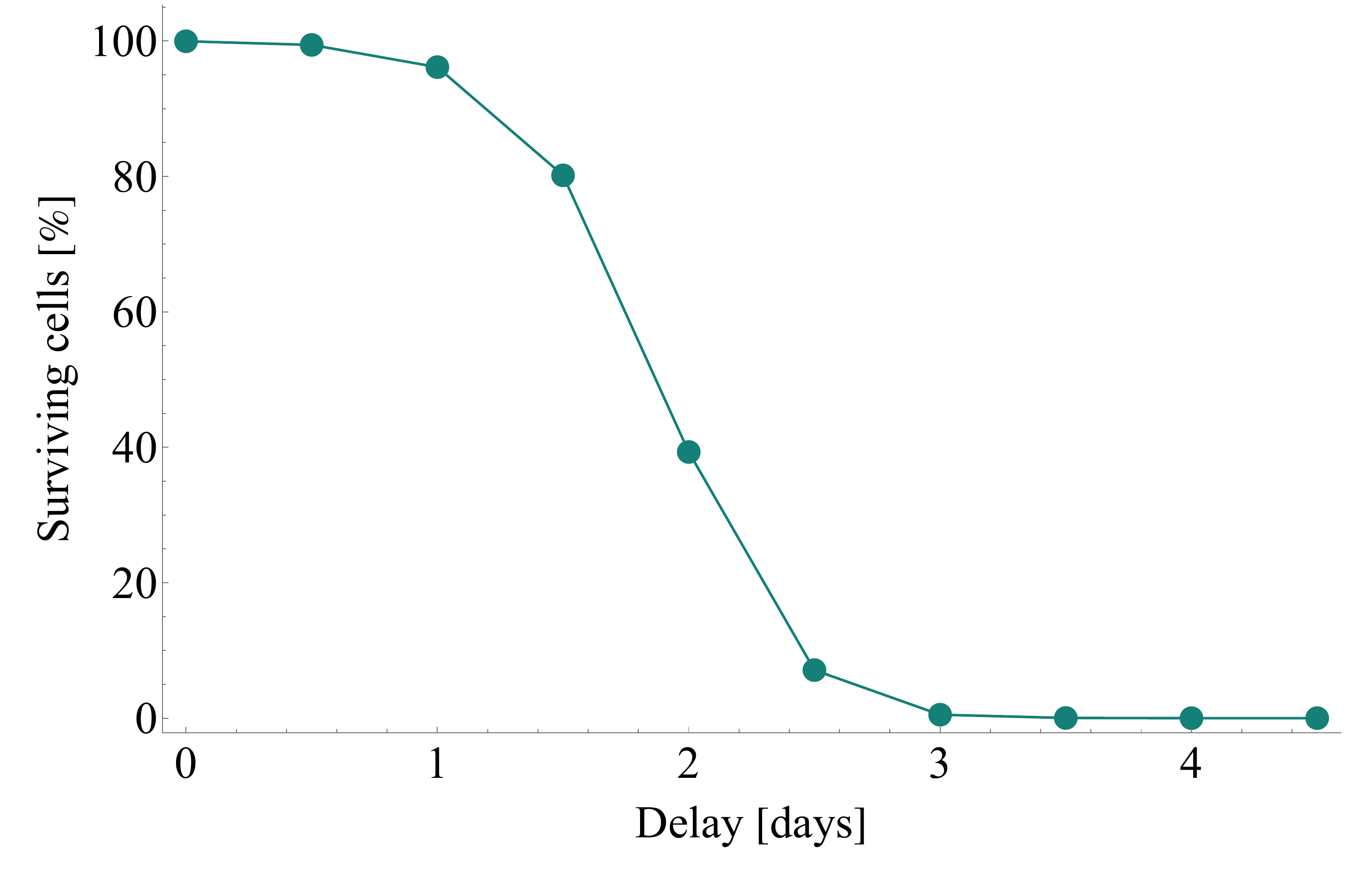} 
        \vspace{-\baselineskip}
        \caption{\empty\label{delayeffectxdiff02-fix}}
    \end{subfigure}
     \begin{subfigure}[b]{0.48\textwidth}
        \includegraphics[width=\textwidth]{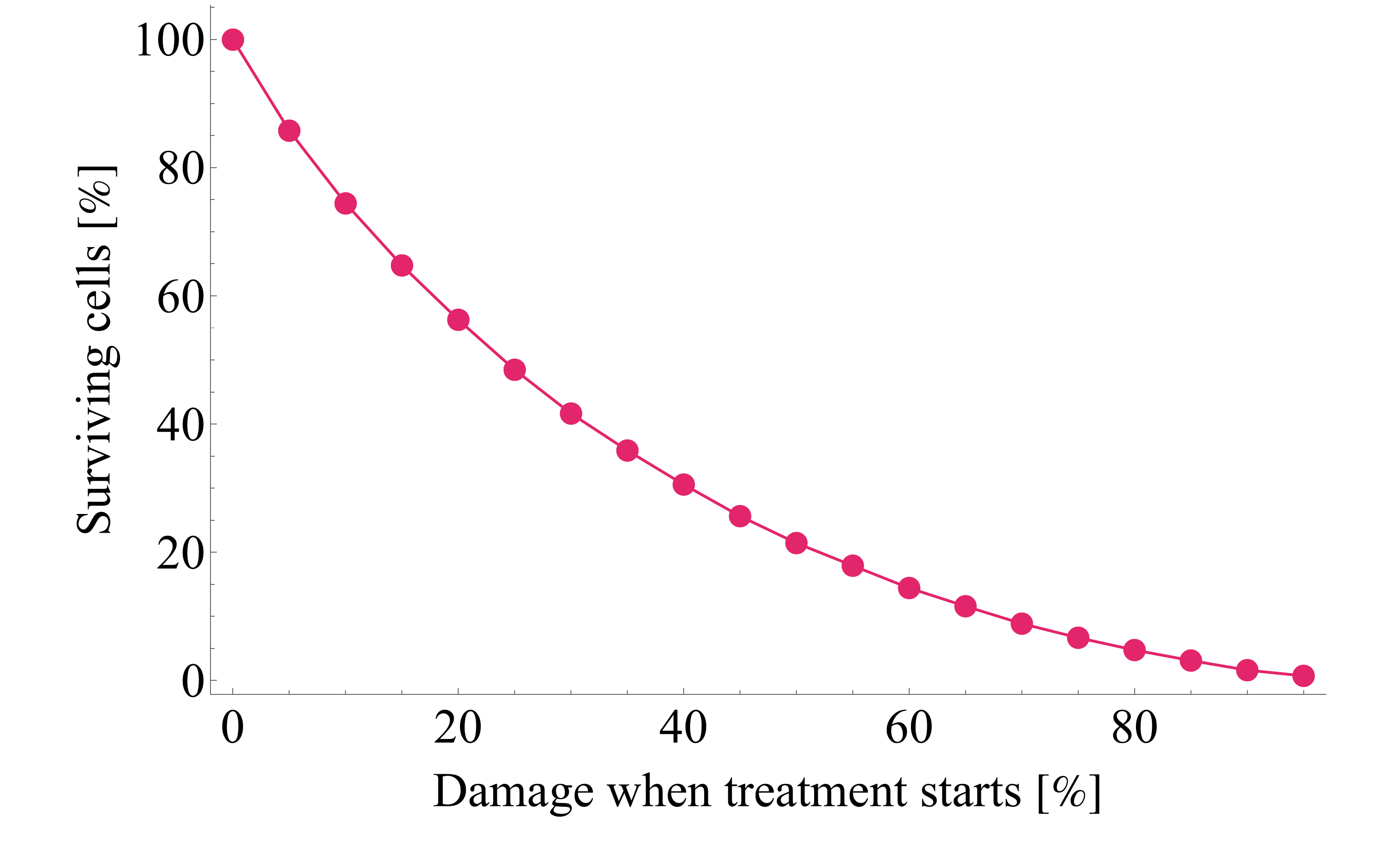}
        \vspace{-\baselineskip}
        \caption{\empty\label{damagedelayxdiff02-damage} }
    \end{subfigure}
    \caption{The damaging effect of treatment delay in two different approaches. Both subfigures illustrate the ratio of remaining uninfected target cells -- the substantial difference between the two plots is the quantity measured on the horizontal axes. Subfigure (\textbf{a}) follows time, directly, on its $x$ axes, while graph (\textbf{b}) depicts results w.r.t. initial damage rates. Results were calculated with the same fixed diffusion coefficient as used in \cite{hybrid-PDE-ABM-1}, namely, $D_V = 0.2 \sigma^2 / \text{min}.$}
    \label{delayeffectxdiff02}
\end{figure}

Figure~\ref{delayeffectxdiff02} illustrates the damaging effect of treatment delay from two different viewpoints.

The first one, Figure~\ref{delayeffectxdiff02-fix} considers averted damage for a series of scenarios where each scenario assumes a $12$--hour additional delay compared to the previous one. We highlight the sharp fall in effectiveness after a delay of $1.5$ days: lack of timely Paxlovid-based antiviral intervention proves to be the most costly at this exact time window. For clarity we note that the expression \textit{surviving cells} refers to the fraction of initially susceptible cells that has not become infected by the end of the observation period.

The main idea of the second approach (Figure~\ref{damagedelayxdiff02-damage}) is to redefine the quantity on the horizontal axis: here, instead of linearly increasing delay times, the $x$ axis follows initial damage rates (i.e. the level of damage that has been done until the moment treatment with Paxlovid is started). In other words, the latter approach depicts the relation between initial damage and averted damage.

\begin{figure}[H]
    \centering
    \begin{subfigure}[c]{0.7\textwidth}
        \includegraphics[width=\textwidth]{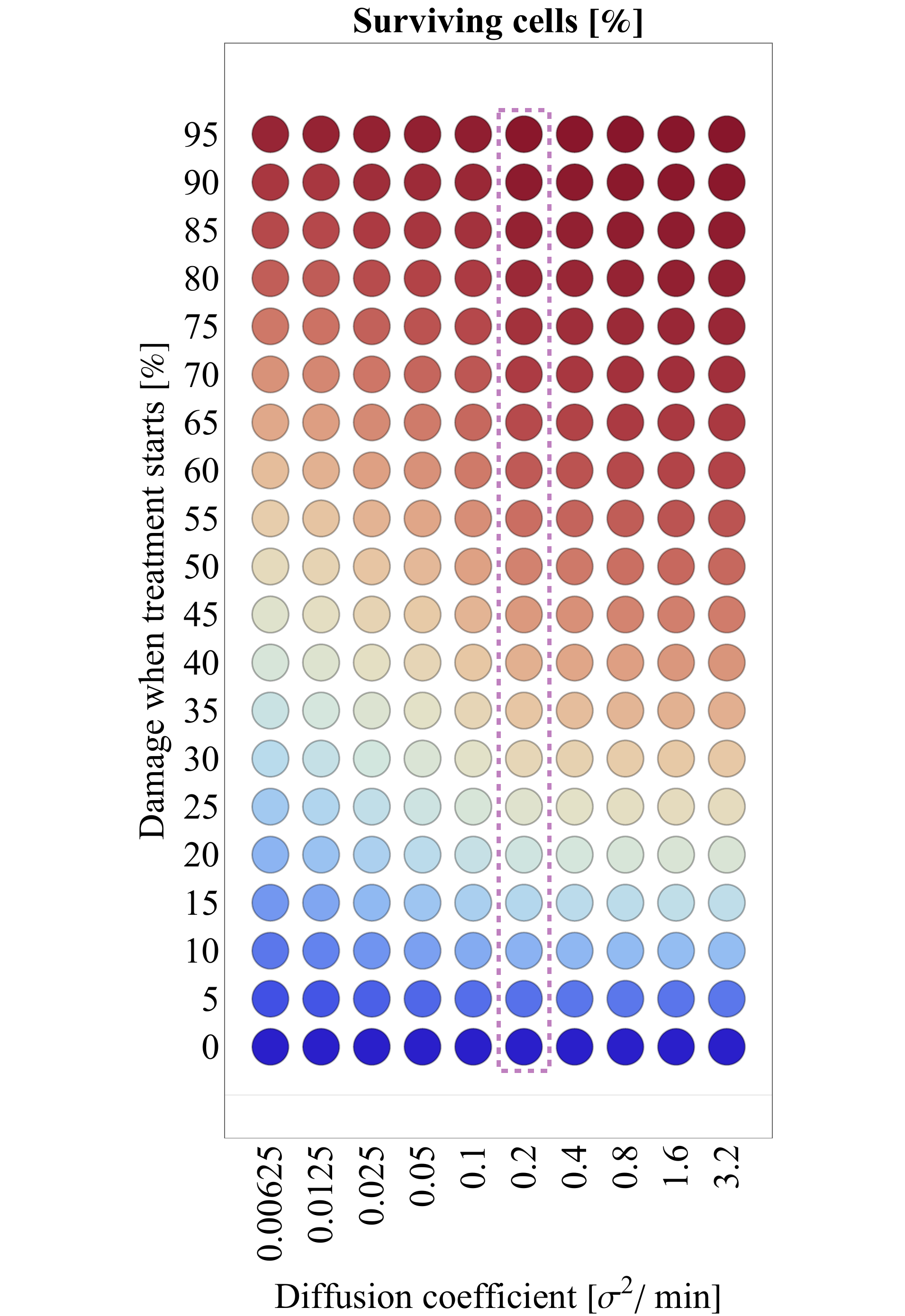}
    \end{subfigure}
     \begin{subfigure}[c]{0.07\textwidth}
        \includegraphics[width=\textwidth]{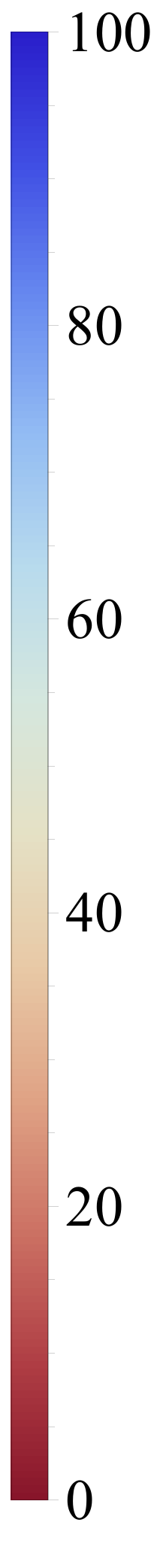}
    \end{subfigure}
    \caption{Interplay between the virus diffusion coefficient (horizontal axis) and tissue damage at the initialization of Paxlovid treatment (vertical axis). The column corresponding to the particular (default) virus diffusion value of $D_V=0.2 \sigma^2 / \text{min}$ (the one used in \cite{hybrid-PDE-ABM-1}) is highlighted with purple.}
    \label{heat}
\end{figure}

Due to the lack of precise clinical data -- and consequently, the relative uncertainty -- regarding the exact diffusion coefficient value of SARS--CoV--2 we explore the respective sensitivity of the results shown in Figure~\ref{damagedelayxdiff02-damage}. Specifically, Figure~\ref{heat}  illustrates the corresponding results in a heatmap for different diffusion values. Compared to the default scenario assuming $D_V = 0.2,$ the outcomes do not change substantially for even significantly higher $D_V$ values, however, there is a clear pattern suggesting that infection outcomes are expected to be more favorable if the diffusion coefficient is several magnitudes lower.

\begin{figure}[H]
    \centering
    \includegraphics[width=0.6\textwidth]{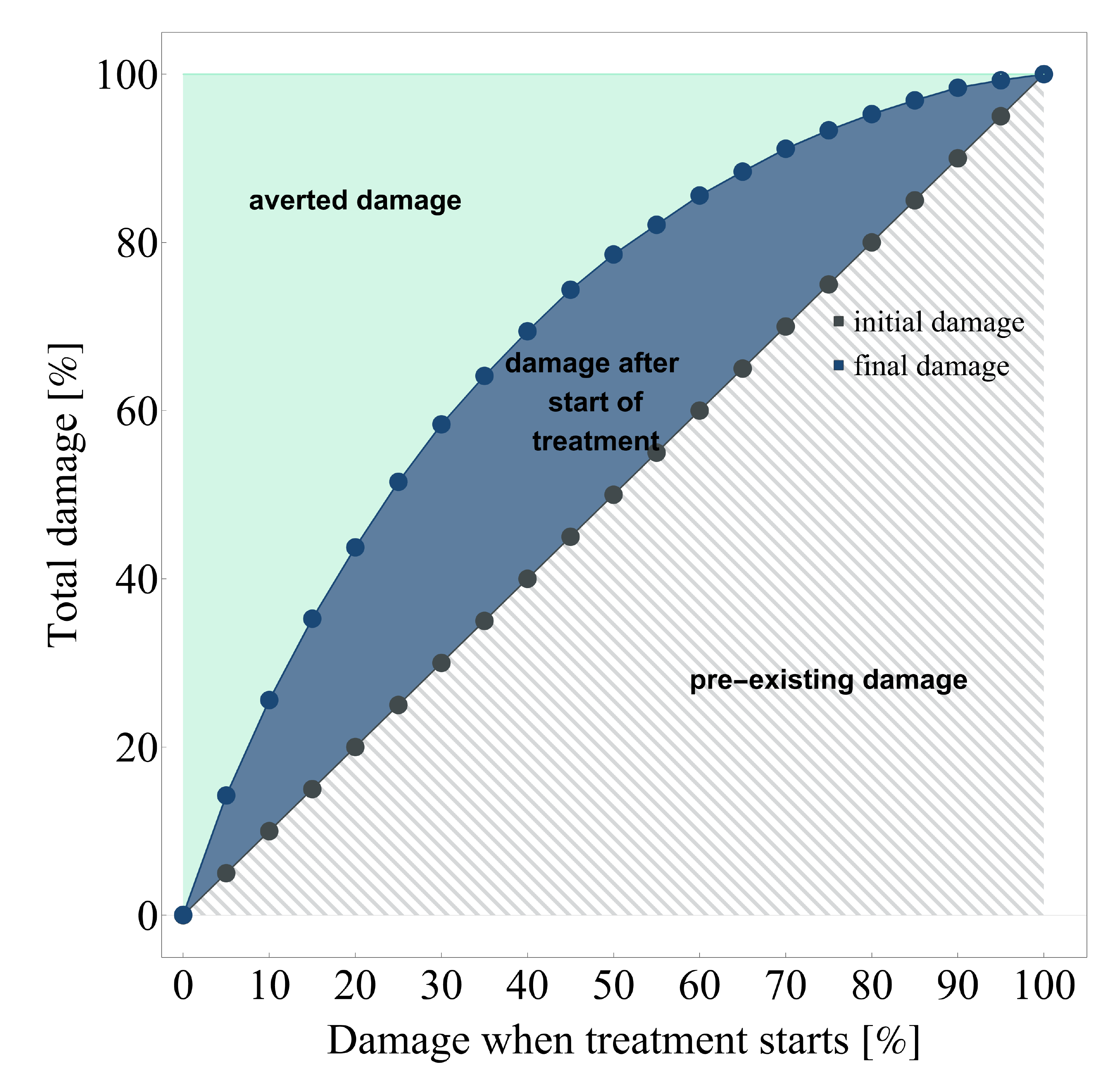}
    \caption{The visualization of averted damage as a result of Paxlovid treatment. The quantity on the horizontal axis (and the $x=y$ line itself) represents the level of cell culture damage suffered until Paxlovid treatment begins, while data points depicted in blue show the unavoidable further damage that occurs after therapy commences. The shaded areas are a precise visual representation of \textit{initial damage} (dashed), \textit{unavoidable post-intervention damage} (blue), and \textit{averted damage} (green). Evidently, the light green area represents those healthily functioning epithelial lung cells that were ultimately saved by Paxlovid.}
    \label{the.Plot}
\end{figure}

Finally, we visualize the potency of Paxlovid in Figure~\ref{the.Plot}: in this graph we principally approach damage rates as areas. While this image is similar to Figure~\ref{damagedelayxdiff02-damage} in that the horizontal axis corresponds to initial damage, Figure~\ref{the.Plot} is ultimately structured differently. It distinguishes three types of damages and represents them as two-dimensional volumes -- namely, we consider \textit{initial damage}, \textit{damage after treatment initialization}, and \textit{averted damage}.

Naturally, the area between the $x=y$ line -- depicted in (dotted) gray -- and the horizontal axis corresponds to the level of tissue damage suffered until the moment of Paxlovid-based intervention, i.e. \textit{initial damage}.

As our next step, we visualize the unavoidable damage that occurs after intervention begins: the (dotted) curve depicted with blue shows the further damage that takes place even after the patient starts taking Paxlovid. Evidently, the area between the gray line and the blue curve is the visual representation of \textit{damage after treatment initialization}. This rate of damage is especially high when soaring virus concentration values are combined with a significant fraction of susceptible target cells at the initialization time of Paxlovid treatment. The latter is explained simply by nirmatrelvir's mechanism of action: nirmatrelvir does effectively block virus production in infected cells, but it can not prevent target cells from getting infected, which is also apparent in the figure itself.

The third category, \textit{averted damage} emerges in Figure~\ref{the.Plot} as the area between the blue curve and the horizontal line framing the graph from above (the latter naturally corresponds to the scenario where no medical intervention happens and full-scale damage takes place after $5$ days). This shaded, light green area is the visual equivalent of the damage that is averted as a result of Paxlovid treatment, or in other words, the epithelial lung cells that are saved by this new $\text{M}^\text{pro}$ inhibitor. Similarly to numerous other antiviral drugs (targeting a large variety of viruses), the principle of 'the sooner the better' proves to hold in this case, too: if intervention happens right at the beginning, almost the entire cell population can be saved by Paxlovid in case of a SARS--CoV--2 infection.

\section{Discussion}

Even with worldwide vaccination programmes, SARS--CoV--2 and its newly emerging variants represent an unprecedented global challenge. Consequently, new alternative treatment options are still very much needed. This paper yields a mathematical, computation-based evaluation of one of the most promising SARS--CoV--2 inhibitors to date, Paxlovid. We implemented and carefully calibrated a multiscale mathematical framework to serve as a small \textit{in silico} laboratory where the basic features of Paxlovid can be replicated, explained, and further investigated. Our calculations correspond to clinical expectations remarkably well: we successully replicated the outcome of a real-life \textit{in vitro} experiment in the simulated context of our model, moreover, both the sufficiency and the necessity of Paxlovid's two main components were verified by our computations for a simplified \textit{in vivo} case. To further improve Paxlovid's assessment, we generated a heat map investigating the results' sensitivity to the inherently vaguely specified virus diffusion coefficient. Despite the mathematical model's necessary simplifications and the short scope of this case study we were able to visualize and verify the importance of prophylactic interventions, moreover, we identified the specific time window where delaying treatment initiation proves to be most costly.

As for directions of future work, we highlight that such hybrid models and computational frameworks hold a great deal of promise with applications such as supporting clinical trials by means of in silico experiments. Computation-based evaluation and simulation of therapies not only can enhance optimization of treatments, but a further development of this technology could also serve to reduce the need for animal testing in the future.

\section*{Acknowledgement}
Authors were supported by TKP2021-NVA-09 and the National Research, Development and Innovation
Fund of Hungary grants FK 138924 (FB),  KKP 129877 (NJ,SM,RH), FK 124016 (GR). In addition, FB was also supported by \'UNKP-21-5 and the 
Bolyai Scholarship of the 
Hungarian Academy of Sciences.

\end{document}